\documentclass[aps,prd,
superscriptaddress,
bibnotes,
amsmath,
amssymb,
nofootinbib,
floatfix]{revtex4-2}
\usepackage[utf8]{inputenc}
\usepackage{hyperref}
\usepackage{xcolor}
\usepackage{booktabs}
\usepackage{graphicx}

\usepackage[T1]{fontenc}

\usepackage{setspace}
\usepackage{xcolor}
\usepackage{physics}
\usepackage{subfig}
\usepackage[font=small]{caption}

\newcommand\mib{Dipartimento di Fisica, Universit\`a degli Studi di Milano-Bicocca, Piazza della Scienza 3, I-20126 Milano, Italy}
\newcommand\infn{INFN, Sezione di Milano-Bicocca, Piazza della Scienza 3, I-20126 Milano, Italy}

\def\obs{\mathcal O}

\def\obsbar{\overline{\vphantom{(}\obs}}
\def\dobsbar{\delta \overline{\vphantom{(}\obs}}

\def\Tr{\mathrm{Tr}\,}

\def\Gbar{\overline{\vphantom{(}\Gamma}}
\def\Cbar{\overline{\vphantom{(}C}}
\def\bnd{\mathrm{bnd}}

\def\taueff{\tau_{\mathrm{eff},\alpha}}

\def\Cbar{\overline{C}}

\def\CPN{\mathbb{CP}^{M-1}}

\renewcommand{\vec}[0]{\boldsymbol}

\begin{document}

\title{Bounding statistical errors in lattice field theory simulations}
\author{M.~Bruno\email{mattia.bruno@unimib.it}}\affiliation{\mib}\affiliation{\infn}
\author{G.~Morandi\email{gabriele.morandi@unimib.it}}\affiliation{\mib}\affiliation{\infn}

\begin{abstract}
    Simulations of strongly interacting lattice field theories are typically performed using Markov chain Monte Carlo algorithms. Therefore estimators of statistical errors must incorporate the effect of autocorrelations by integrating the corresponding autocorrelation function. Since in practical calculations its integral is truncated to a finite window, in this work we propose a stopping criterion based on upper and lower bounds of the autocorrelation function. We examine its application to both traditional Monte Carlo analysis and the recently introduced master-field approach. By leveraging both bounds, we introduce an automatic windowing procedure which we test on numerical simulations of a few simplified toy models.
\end{abstract}
\maketitle

\section{Introduction}

Modern Lattice QCD calculations have recently reached a new high level of accuracy and precision, for example, in the context of the muon anomalous magnetic moment, motivated by its strong phenomenological implications (see for instance Refs.~\cite{Aoyama:2020ynm, Aliberti:2025beg}). While often the most difficult part consists in providing a reliable estimate of the so-called systematic effects, which in typical lattice calculations amount to discretization, finite-volume and quark-mass mistuning errors, controlling statistical errors is in some cases a point of similar importance. Consider for instance simulations with a low number of representative field configurations, e.g. in the presence of strong autocorrelations at finer lattice spacings.

Expectation values of correlation functions are calculated as statistical averages from a Monte Carlo sampling of the path integral (rotated to the Euclidean metric). Since Markov chains are typically used, and simulations are often very costly from the numerical point of view, correlations along the so-called Monte Carlo direction are non-negligible and have to be taken into account, to avoid underestimating the errors. More importantly, as the lattice spacing becomes finer, algorithms suffer from critical slowing down, increasing the problem's difficulty~\cite{DelDebbio:2004xh}.
Errors are defined from the area of the so-called autocorrelation function~\cite{Madras:1988ei} and in practical calculations a windowing procedure for its truncation is required. This approach to error analysis is typically called the $\Gamma$-method and the works by Madras and Sokal~\cite{Madras:1988ei} and by U.~Wolff~\cite{Wolff:2003sm} (and extensions~\cite{Schaefer:2010hu}) have become standard references, with several software libraries implementing them, see e.g~\cite{Joswig:2022qfe, Ramos:2018vgu, mattia_bruno_2024_14170562}.
Physically interesting quantities are obtained from mathematical manipulations of the primary observables, and recently further progress in the context of fitting has been achieved for cases where the covariance matrix is poorly known, e.g. due to large autocorrelations~\cite{Bruno:2022mfy, Christ:2024nxz, Kelly:2019wfj}. Once calculated, the errors of derived observables may be estimated from their corresponding autocorrelation functions, obtained by properly projecting the fluctuations of the primary observables (using for example linear error propagation)~\cite{Wolff:2003sm}. Our work is centered around the autocorrelation function and its area, but we note that, as an alternative, one can also consider resampling techniques~\cite{efron1982jackknife, Quenouille1949, Tukey1958, Efron:1979bxm} combined with binning strategies to incorporate the effect of autocorrelations, a subject that we do not cover here.   

Recently, a different paradigm for the definition of the error has been introduced. It is based on the notion of stochastic locality~\cite{Luscher:2017cjh} and in this case estimators are constructed from a few configurations of a large-volume simulation, capable to accommodate a sufficiently large number of fluctuations of the studied observables (for some recent studies see for instance~\cite{Giusti:2018cmp, Fritzsch:2021klm, Bruno:2023vhs, RBC:2023pvn}). Statistical averages are replaced by translational averages, and consequently error estimators are built from correlation functions whose behavior is dictated by the properties of the theory, rather than the algorithm, a fact that we try to exploit in our study.
Throughout this work we refer to this method as the \emph{master-field} approach, without necessarily implying the concept of a single configuration on an extremely large volume, but rather the notion of leveraging stochastic locality for the error estimate, even when several configurations are available. 

Evidently statistical error estimators remain an important evolving topic within the lattice field theory community and the purpose of this work is to continue along this path, by presenting an alternative methodology for estimating errors reliably, both in traditional Monte Carlo (MC) and one-dimensional master-field (MF) analysis. The latter may be realized on a large volume by considering only a single direction, or on a very long lattice~\cite{Bruno:2022ljo}.
The key concept that we explore in this manuscript is the possibility to define upper and lower bounds for the autocorrelation function to be used in a windowing procedure (later denoted as bounding method) for the error estimation. For the formulation of these bounds we have been inspired by similar considerations applied to physically relevant problems, such as the calculation of the Hadronic-Vacuum-Polarization contribution to the muon $(g-2)$~\cite{Lehner:2016BNL, Borsanyi:2016lpl}, where they have been used successfully.

Our manuscript is organized as follows: in Section~\ref{sec:Bounding} the bounding method is described and discussed, together with possible improvements and its applicability, while in Section~\ref{sec:Results} we report a few numerical examples where we apply the method to both synthetic and real data, before concluding.

\section{The Bounding Method}\label{sec:Bounding}

We begin by introducing a unified notation that will help us describing both the MC and MF analysis, since they share several important similarities.
We consider $N$ estimators $\obs_\alpha(t)$ of several observables $\obs_\alpha$ obtained from a generic stochastic process. The basic idea is that the parameter $t$ may label either Monte Carlo time or physical time for one-dimensional master-field analysis, with $N$ adjusted accordingly to represent the number of available points\footnote{We call ``time'' the direction where the master-field analysis is performed, but in Euclidean space this is evidently a general approach applicable to any direction.}. For the latter, $t$ is given in units of the lattice spacing to keep a uniform notation, and in general we further consider $t$ to be contiguous and regularly distributed\footnote{The interested reader may check Eq.~(3.18) in Ref.~\cite{Bruno:2023vhs} for an irregular distribution.}.
The central value of $\obs_\alpha$ is given by
\begin{equation}
    \obsbar_\alpha = \frac{1}{N} \sum_{t} \obs_\alpha(t) \,,
\end{equation}
and we denote with $\expval{\obs_\alpha}$ the true expectation value, given by averaging $\obsbar_\alpha$ over an infinite number of repetitions (replica) of the stochastic process. 
For later convenience we introduce the unbiased fluctuations
\begin{equation}\label{eq:fluctuations}
    \delta \obs_\alpha(t) \equiv \obs_\alpha(t) - \expval{\obs_\alpha} \,, \quad \text{and} \quad \delta \obsbar_\alpha(t) \equiv \obs_\alpha(t) - \obsbar_\alpha \,,
\end{equation}
and the central object of our work, the autocorrelation function
\begin{equation}
    \Gamma_{\alpha\beta}(t-t') = \Gamma_{\alpha\beta}(|t-t'|) \equiv \expval{\delta \obs_\alpha(t) \, \delta \obs_\beta(t')} \,.
    \label{eq:gamma_def}
\end{equation}
The most important working hypothesis, beyond invariance under translations of the index $t$, i.e. $\expval{\obs_\alpha(t)} = \expval{\obs_\alpha(0)} = \expval{\obs_\alpha}$, is that the autocorrelation function has an exponentially decreasing behavior, namely that
\begin{equation}
    \Gamma_{\alpha\beta}(t)= \sum_n c^n_{\alpha\beta} \, e^{-|t|/\tau_n} \,, \qquad \tau_n > \tau_{n+1} \,, \; \forall n \,,
    \label{eq:gamma_exp}
\end{equation}
where $\tau_0$ dominates the asymptotic decay and the coefficients $c^n_{\alpha\beta}$ are observable dependent and satisfy $c_{\alpha\alpha}^n>0$.
Note that while for the MC analysis the notation with the modes $\tau_n$ is fairly standard, for the MF one we refer to such modes as the ``energies'' $E_n = 1/\tau_n$, since in this case they correspond to the spectrum of the theory. Below we exploit the consequences of the assumptions in Eq.~\eqref{eq:gamma_exp} and we comment on the concrete cases when these are realized in subsection~\ref{subsec:applicability}.

The error of $\obsbar_\alpha$ is given by the standard formula (repeated indices below are not summed)
\begin{equation}
    \sigma^2_{\obs_\alpha} = \expval{\left[ \obsbar_\alpha - \expval{\obs_\alpha} \right]^2} = \frac{1}{N^2} \sum_{t,t'} \Gamma_{\alpha\alpha}(t-t') \simeq \frac{C_{\alpha\alpha}}{N} \,,
    \label{eq:sigma}
\end{equation}
with
\begin{equation}
    C_{\alpha\beta} = \sum_{t=-\infty}^\infty \Gamma_{\alpha\beta}(t) \,.
\end{equation}
The right hand side of Eq.~\eqref{eq:sigma} receives sub-leading corrections that vanish for large $N$ and that we neglect. More specifically, given the exponential decay in Eq.~\eqref{eq:gamma_exp} and assuming the dominance of $\tau_0$, these sub-leading corrections may be estimated provided that the boundary conditions of the sequence of estimators $\obs_\alpha(t)$ are also considered. For the periodic case (e.g. MF simulations with periodic boundaries) they scale as $O\!\left( e^{-N/\tau_0} \right)$~\cite{Luscher:2017cjh, Bruno:2023vhs}, whereas for the open case with  $O(\tau_0/N)$~\cite{Wolff:2003sm}. For completeness we introduce also the integrated autocorrelation time
\begin{equation}\label{eq:tau_int}
    \tau_{\mathrm{int},\alpha} = \frac12 \frac{C_{\alpha\alpha}}{\Gamma_{\alpha\alpha}(0)} \,, 
\end{equation}
and briefly review the case of derived observables, namely $f(\{\obs_\alpha\})$, with $f$ a generic function. Following Ref.~\cite{Wolff:2003sm}, the fluctuations of the estimator $\overline f = f(\{\obsbar_\alpha\})$, are calculated from the first derivatives of $f$ according to
\begin{equation}
    \delta \overline f(t) \equiv \sum_\alpha f_\alpha \dobsbar_\alpha(t) \,, \quad \text{with} \quad 
    f_\alpha \equiv \frac{\partial f}{\partial \obs_\alpha} \Big|_{\obs_\alpha = \obsbar_\alpha} \,.
\end{equation}
The corresponding autocorrelation function is defined by 
\begin{equation}
    \Gamma_f(t) = \sum_{\alpha\beta} f_\alpha \, \Gamma_{\alpha\beta}(t) \, f_\beta 
\end{equation}
and admits a decomposition as in Eq.~\eqref{eq:gamma_exp}, with the coefficients $c_{\alpha\beta}^n$ replaced by $c_f^n = \sum_{\alpha\beta} f_\alpha c_{\alpha\beta}^n f_\beta$.

A good practical estimator for the autocorrelation function is
\begin{equation}\label{eq:estimator_G}
    \Gbar_{\alpha\beta}(t) = \frac{1}{\mathcal{N}(t)} \sum_{t'} \dobsbar_\alpha(t+t') \, \dobsbar_\beta(t') \,,
\end{equation}
where the normalization $\mathcal{N}(t)$ takes into account the fact that the sequence of measurements $\obs_\alpha(t)$ may be subject to periodic ($\mathcal{N}(t)=N$) or open boundary conditions ($\mathcal{N}(t)=N-t$), depending on whether one is interested in MF or MC analysis respectively. It is well known~\cite{Wolff:2003sm,Bruno:2023vhs} that the estimator above has a small bias
\begin{equation}
    \expval{\Gbar_{\alpha\beta}(t)} - \Gamma_{\alpha\beta}(t) \simeq - \frac{C_{\alpha\beta}}{N}
\end{equation}
that we neglect in the following. 
The error of $\Gbar_{\alpha\beta}(t)$ may be calculated numerically, up to a normalization by $1/N$, by summing over $t''$ the following expression
\begin{equation}\label{eq:err_gamma}
    \frac{1}{\mathcal N(t+t'')} \sum_{t'} \big[ \dobsbar_\alpha(t+t') \dobsbar_\beta(t') - \Gbar_{\alpha\beta}(t) \big] \big[ \dobsbar_\alpha(t+t'+t'') \dobsbar_\beta(t'+t'') - \Gbar_{\alpha\beta}(t) \big]
\end{equation}
for fixed $\alpha,\beta$ and $t$ (see also Ref~\cite{Yamamoto:2025imx}). Alternatively one may use the analytic approximated formula reported in Ref.~\cite{Luscher:2004pav}, Appendix E. For the results reported in the next Section we adopt the direct numerical estimate above calculated with the library \texttt{pyobs}~\cite{mattia_bruno_2024_14170562}.

Finally, to estimate the variance from $\Gbar_{\alpha\beta}(t)$ we adopt
\begin{equation}\label{eq:cbar_W}
    \Cbar_{\alpha\beta}(W) = \sum_{|t| \leq W} \Gbar_{\alpha\beta}(t) \,,
\end{equation}
where we have additionally truncated the sum to a finite window $W$ due to the finite statistics in practical simulations\footnote{In fact setting $W \simeq N$ would lead to an estimator $\Cbar_{\alpha\beta}$ whose variance would not converge to zero in the infinite statistics limit, as noted by Madras and Sokal in Ref.~\cite{Madras:1988ei}.}.
Similarly, for the integrated autocorrelation time one can replace $C_{\alpha\beta}$ with $\Cbar_{\alpha\beta}(W)$ in Eq.~\eqref{eq:tau_int}.

\subsection{Upper and lower bounds}

Now we focus explicitly on errors thereby restricting ourselves to the diagonal entries of the covariance matrix. Following a similar approach adopted in the calculation of the Hadronic-Vacuum-Polarization contribution to the muon anomaly~\cite{Lehner:2016BNL, Borsanyi:2016lpl, RBC:2018dos, Bruno:2019nzm}, by leveraging the positivity of the coefficients $c_{\alpha\alpha}^n$ in Eq.~\eqref{eq:gamma_exp} we introduce the strict lower and upper bounds
\begin{equation}
    \Gamma_{\bnd,\alpha}(t|W, \taueff^{W}) \leq \Gamma_{\alpha\alpha}(t) \leq \Gamma_{\bnd,\alpha}(t|W, \tau_0) \,,
    \label{eq:bounds}
\end{equation}
with
\begin{equation}\label{eq:G_bound}
    \Gamma_{\bnd,\alpha}(t | W, \tau) = 
    \begin{cases}
        \Gamma_{\alpha\alpha}(t) \,, & \text{if} \quad |t| < W \\
        \Gamma_{\alpha\alpha}(W) e^{-(|t|-W)/\tau} \,, & \text{if} \quad |t| \geq W \,.
    \end{cases}
\end{equation}
The effective mode $\taueff^W$ is defined from the logarithmic derivative of the autocorrelation function
\begin{equation}
    1 / \taueff^{W} = 
    - \widetilde \partial_t \log \Gamma_{\alpha\alpha}(t) \big|_{t=W} \,,
    \label{eq:tau_eff}
\end{equation}
where by $\widetilde \partial_t$ we denote an approximation of the derivative based on the available discrete values of $t$. In practice this may be taken as the (inverse of the) usual effective-mass formula $\log[\Gamma_{\alpha\alpha}(t)/\Gamma_{\alpha\alpha}(t+1)]$ used for correlation functions in lattice simulations. 
Since $\Gamma_{\alpha\alpha}(t)$ is a log-convex function\footnote{
To prove the log-convexity focusing on $t > W$ is sufficient. To simplify the derivation we drop the indices ${}_{\alpha\alpha}$, $\Gamma(t) \equiv \Gamma_{\alpha\alpha}(t)$, $c^n \equiv c_{\alpha \alpha}^n$. 
By inserting Eq.~\eqref{eq:tau_eff} inside Eq.~\eqref{eq:G_bound}, we find for the lower bound $\Gamma (W) e^{(t-W)\partial_t \log \Gamma(t) |_{t=W}} \leq \Gamma(t)$.
Using $\gamma(t) \equiv \log \Gamma(t)$ we arrive at the inequality
$$
    (t-W)\partial_t \gamma(t) \big|_{t=W} \leq \gamma(t) - \gamma(W) \,,
$$
which holds if and only if $\gamma(t)$ is convex, i.e. $\partial^2_t \gamma(t) \geq 0$. The latter follows straightforwardly from Eq.~\eqref{eq:gamma_exp}
$$
    [\Gamma(t)]^2 \, \partial^2_t \gamma(t) = \sum_{n,m} c^n c^m e^{-t(1/\tau_n + 1/\tau_m)} \left( \frac{1}{\tau^2_n} - \frac{1}{\tau_n \tau_m} \right) = \frac{1}{2} \sum_{n,m} c^n c^m e^{-t(1/\tau_n + 1/\tau_m)} \left( \frac{1}{\tau_n} - \frac{1}{\tau_m} \right)^2 \geq 0 \,.
$$
} 
and its derivative is monotonically increasing, $\Gamma_{\bnd,\alpha}(t|W,\taueff^W)$ is a lower bound for $|t| > W$.

On the contrary multiplying $\Gamma_{\alpha\alpha}(W)$ by $e^{-(|t|-W)/\tau_0}$ amounts to replacing all the modes $1/\tau_n \to 1/\tau_0$ for $|t|>W$ automatically implying a slower fall-off of $\Gamma_{\bnd,\alpha}(t|W,\tau_0)$ w.r.t. $\Gamma_{\alpha\alpha}(t)$. 
In practical lattice simulations the estimator of the autocorrelation function is itself affected by errors and subject to statistical fluctuations. As a consequence the effective mode $\taueff^W$, typically obtained from two consecutive time slices, quickly becomes unstable for large values of $t$. In such cases one may fix $\taueff^W$ to the values obtained from earlier time slices, without spoiling the validity of the bound, since $\taueff^W < \taueff^{W+1}$. 

For the calculation of the error we split the \emph{integral} of the autocorrelation function into two contributions
\begin{equation}
    C_{\alpha\alpha} = \sum_{t=-\infty}^\infty \Gamma_{\alpha\alpha}(t) = C_{\alpha\alpha}(W) + 
    \sum_{|t|>W} \, \, \Gamma_{\alpha\alpha}(t) \,, \quad
    C_{\alpha\alpha}(W) \equiv \sum_{|t| \leq W} \, \Gamma_{\alpha\alpha}(t) \,.
    \label{eq:sum_w_gamma}
\end{equation}
and consequently calculate the integral of the bounds in Eq.~\eqref{eq:G_bound} as 
\begin{equation}\label{eq:int_bounds}
    C_{\mathrm{bnd},\alpha} (W,\tau) \equiv \sum_{t=-\infty}^{\infty} \Gamma_{\mathrm{bnd},\alpha}(t|W,\tau) = C_{\alpha\alpha}(W) + 2 \Gamma_{\alpha\alpha}(W) \frac{e^{-1/\tau}}{1-e^{-1/\tau}} \,.
\end{equation}

Starting from Eq.~\eqref{eq:bounds}, the systematic error from the sum over $|t|>W$ in Eq.~\eqref{eq:sum_w_gamma}
is consequently bounded from above and below
\begin{equation}\label{eq:sum_bounds}
    \sum_{|t|>W}  \, \Gamma_{\bnd,\alpha}(t|W,\taueff^{W}) 
    \leq \sum_{|t|>W} \, \Gamma_{\alpha\alpha}(t) \leq
    \sum_{|t|>W} \, \Gamma_{\bnd,\alpha}(t|W,\tau_0) \,,
\end{equation}
allowing us to formulate an automatic windowing procedure to calculate $W$ such that the systematic error introduced by the truncation at $W$ is in good balance with the statistical error of $C_{\alpha\alpha}(W)$, denoted by $\sigma_{\mathrm{stat},\alpha}(W)$.
Similarly to Refs.~\cite{Lehner:2016BNL, Borsanyi:2016lpl}, the systematic effect of the truncation is obtained from the difference of the integrals of the upper and lower bounds in Eq.~\eqref{eq:int_bounds} (strictly positive since $\taueff^W \leq \tau_0$ $\forall \, W$)
\begin{equation}\label{eq:sys_error}
    \sigma_{\mathrm{sys,\alpha}}(W) = 
    \sum_{|t|>W} \, \big[\Gamma_{\bnd,\alpha}(t|W, \tau_0) - 
    \Gamma_{\bnd,\alpha}(t|W, \taueff^{W}) \big] = 2 \Gamma_{\alpha\alpha}(W) \frac{e^{-1/\tau_0} - e^{-1/\taueff^W}}{(1-e^{-1/\taueff^W})(1-e^{-1/\tau_0})} \,,
\end{equation}
and it may be used to define an optimal summation window from the solution of the equation $\sigma_{\mathrm{stat},\alpha}(W) = M \sigma_{\mathrm{sys},\alpha}(W)$ with $M$ a positive user-defined number larger than 1, but conceivably $M \geq 2$. In the numerical results presented in the next Section, $M$ is always set to 2.
The error of the error may be estimated from data or taken from analytic approximations. For example, for the MC analysis by adopting the formula from Ref.~\cite{Wolff:2003sm} one can re-write the stopping criterion as
\begin{equation}\label{eq:stopping_criterium}
    \frac{\sigma_{\mathrm{stat},\alpha}(W)}{C_{\alpha\alpha}(W)} = \sqrt{\frac{2 (2W+1)}{N}} =
    M \frac{\sigma_{\mathrm{sys},\alpha}(W)}{C_{\alpha\alpha}(W)} \,.
\end{equation}
The criterion above may be applied also to the one-dimensional MF analysis~\cite{Bruno:2023vhs}, by replacing $N$ with the number of available points. Since in practice the bounding method is always applied to estimators, which above we indicated with overline bars, to keep a light and readable notation we will use $\Gamma_{\bnd}$ (and $C_{\bnd}$) throughout the rest of the manuscript, even if it is calculated from $\Gbar$.

A few considerations follow.
Firstly, at asymptotically large times where only a single exponential contributes to $\Gamma_{\alpha\alpha}(t)$ the two bounds become identical, meaning that, in practice, they may easily saturate even when autocorrelations are still statistically resolvable from zero (see Figs.~\ref{fig:Bounding_MC_synthetic} and \ref{fig:Bounding_MC_synthetic_2}) and the corresponding optimal window may be interpreted as the beginning of single mode dominance. Secondly, when the autocorrelation function falls off very rapidly, one may simply adopt a fixed summation window obtained from visual inspection. Cases where the method proposed here, or the more standard Wolff's approach and its variants~\cite{Schaefer:2010hu}, are relevant is in the presence of large autocorrelations. 
Finally, we remark that the positivity of $\Gamma_{\alpha\alpha}(t)$ is central to the definition of these bounds restricting the applicability of the bounding method to the diagonal entries of the covariance matrix, i.e. to the errors of the observables. A method for the off-diagonal entries $C_{\alpha\beta}$ remains highly desirable. For the error of derived observables the bounding method can be applied straightforwardly, thanks to the positivity of the coefficients $c_f$ (which follows from Eq.~\eqref{eq:gamma_spec_dec}).

\subsection{Applicability}\label{subsec:applicability}

At this point the two fundamental assumptions that entered in our derivation have been the invariance under translations along the direction given by the coordinate $t$ and Eq.~\eqref{eq:gamma_exp}. Under the appropriate conditions, they are both satisfied for one-dimensional Master-Field and also for Monte Carlo analysis, which we discuss below.\\

\textit{Monte Carlo analysis} --- Provided that the algorithm chosen to generate the Markov chain of field configurations satisfies detailed balance and ergodicity, one can demonstrate that the associated transition probability possesses bounded eigenvalues. However $\Gamma_{\alpha\beta}(t)$ obeys a representation as in Eq.~\eqref{eq:gamma_exp} only when $t$ is given in units of an even number of Markov steps (see Appendix~\ref{app:gamma}). The proposed bounding method is therefore applicable in these cases. As observed in Ref.~\cite{Schaefer:2010hu} further selection rules may be relevant in separating the parity even and odd sectors, and in our numerical examples below we focus on the former.

While the lower bound is uniquely determined from data, for the upper bound an a-priori knowledge of $\tau_0$ is needed. In the analysis of a single observable, a possible approach to this problem is to develop a consistent iterative procedure. 
In fact, starting from the right-most inequality in Eq.~\eqref{eq:sum_bounds}, it is possible to show, with little algebra, that 
\begin{equation}\label{eq:tau_hat}
    \widehat{\tau}_0(W') = \frac{1}{\log \! \left(1 + \frac{\Gamma(W')}{\sum_{t>W'} \Gamma(t)}\right)}
\end{equation}
defines a strict upper bound for $W < W'$. 
Afterwards one could consider a first application of the bounding method with $k \widehat \tau_0(W')$ in the upper bound. By choosing $k > 1$ (e.g. $k=2,3$), the new optimal window $W''$ would be greater than $W'$. The repetition of this operation with $k \widehat \tau_0(W'')$ would ensure a convergence of the procedure within a certain accuracy, implying that one could start even from small values of $W'$. In essence the result obtained from one application of the bounding method is used as an input for the next one. 
However due to the presence of statistical errors, estimates of $\widehat \tau_0$ may become unstable for large values of $W$; moreover the need to effectively truncate the summation in Eq.~\eqref{eq:tau_hat} would introduce an additional assumption. A simpler (but non-strict) iterative procedure, would simply consist in adopting $k$ times the integrated autocorrelation time, with its initial value obtained by inspecting the lower bound. This iterative mechanism may be automatized and interrupted when the optimal window $W$ is stable. Given the dependence on the tunable parameter $k$ and the related assumptions, it is therefore similar to Wolff's approach~\cite{Wolff:2003sm}, where one should always monitor the stability of the error.

A second possible approach consists instead in adopting a variational method, like the GEVP described below, to obtain a reasonable estimate of the slowest mode $\tau_0$. In this case several observables are needed, possibly with a different coupling to the various modes involved and a numerical example is presented in the next Section. Evidently for this strategy to be successful a reasonable separation among the individual modes is necessary, which may not be the case for a generic transition (Markov) matrix. As a third alternative, one may take the longest mode observed from an effective-mass analysis, but we stress that it is nevertheless important to examine several observables to cover a sufficiently large space.

In general it is always advisable to inspect the two bounds, and in those cases when the a priori knowledge of $\tau_0$ is not very accurate, one should monitor the stability of the error under sensible variations of the slowest autocorrelation mode, which for larger values of $\tau_0$ lead to more conservative error estimates approaching the ones of Ref.~\cite{Schaefer:2010hu}. 
Similarly to that work, also here the bounding method relies on the knowledge of the slowest autocorrelation modes, and as a consequence similar considerations apply. However, provided that $\tau_0$ is known to sufficient confidence, relatively to Ref.~\cite{Schaefer:2010hu} our proposal incorporates a lower bound and a stopping criterion based on the saturation of the bounds, leading to tighter but still solid summation windows. In summary, for the Monte Carlo analysis, the introduction of the lower bound constitutes one of the main points of our work, while the iterative schemes proposed above may be subject to further refinements in future work. \\

\textit{Master-field analysis} --- Contrary to the MC case where the details of the algorithm matter, in the master-field scenario the behavior of $\Gamma_{\alpha\beta}(t)$ is completely dictated by the physical properties of the theory. More specifically for theories with a mass gap in Euclidean space-time, the spectral decomposition in Eq.~\eqref{eq:gamma_exp} holds and for QCD the lowest mode is in general given by the pion mass. Instead when the observable has a non-zero expectation value, states with vacuum quantum numbers are expected to dominate at long distances~\cite{Bruno:2023vhs}.

As a consequence, the bounding method proposed before can also be adopted in the case of one-dimensional MF analysis, where the presence of a Hamiltonian and transfer matrix guarantee the validity of Eq.~\eqref{eq:gamma_exp} and where the positivity of the coefficients $c_{\alpha\alpha}^n$ trivially follows from its spectral decomposition\footnote{The bounding method should nevertheless be applied at physical distances larger than the cutoff.}. 
For example, if we consider the expectation value of a local operator such as the plaquette, its autocorrelation function is the typical connected two-point function dominated at long distances by the scalar glueball (in a pure gauge theory).
Therefore a significant advantage in this approach is that the slowest mode is typically known to sufficient accuracy, and since we are interested in using it inside the upper bound, a high-precision determination is not needed, and in practical calculations one can always monitor the stability of the windowing procedure with more conservative choices.

A second advantage of the bounding method is that if the two bounds do not saturate, one may quote the upper bound as an estimate of the error. As noted in Ref.~\cite{Bruno:2023vhs}, Fig.~4, mesonic correlators are particularly challenging in this context, because saturating the integral of the autocorrelation function can itself be difficult. In this case quoting a conservative upper bound may be advantageous.
While this last observation is in spirit equivalent to the upper bound proposed in Ref.~\cite{Schaefer:2010hu}, the substantial difference is given by the fact that in the MF approach one can safely take the mass gap as the slowest mode, while for the MC analysis one typically needs long Markov chains to properly estimate the longest autocorrelation modes.

A few limitations however exist and we briefly mention them here. If one is interested in a MF analysis involving more than one dimension, the corresponding saturation of the autocorrelation function may be studied for each direction separately, using the method proposed here. This however becomes quickly impractical and less tractable. As suggested in Ref.~\cite{Luscher:2017cjh} one should leverage the symmetries of the theory, including invariance under rotations leading to 
\begin{equation}
    \Gamma_{\alpha\alpha}(r) = \sum_{|x|=r} \Gamma_{\alpha\alpha}(x) \,,
\end{equation}
where $x$ represents a multi-dimensional coordinate and $r$ its norm. Assuming that its behavior is dominated by the exchange of a single scalar particle, in four dimensions one finds a Bessel function~\cite{Bruno:2023vhs} which tends to an exponential function only at asymptotically large $r$. Hence extending the formulation of rigorous bounds to these cases requires more care and is deferred to future studies.

Another non-trivial subtlety arising in this type of analysis concerns non-local operators or observables. The former is examined and discussed in the next Section by studying local operators constructed from gauge fields smeared (also) along the MF direction. The latter instead is relevant if one is interested in assessing the error of a two-point correlation function, whose autocorrelation function may not always be related to a proper (four-point) correlator of the theory. For instance, for two-point fermionic observables the autocorrelation function may correspond only to a subset of the Wick contractions defining the corresponding four-point correlator. A proper interpretation of subsets of fermionic diagrams is obtained in the framework of Partially Quenched QCD (PQQCD), an unphysical theory that does not admit a proper Hilbert space and a positive Hamiltonian (see Ref.~\cite{Sharpe:2006pu} for a review on the subject). As a consequence the MF autocorrelation function may not admit a spectral decomposition as in Eq.~\eqref{eq:gamma_exp}, preventing the (strict) usage of the bounding method. However since it is possible to construct a low energy effective field theory of PQQCD~\cite{Sharpe:2006pu}, it is not unconceivable that the bounding method may still work in practice, at least at intermediate and long distances. Given these complications we defer this study to future work.\\

Provided that the dependence of $c_{\alpha\beta}^n$ on $\alpha$ and $\beta$, at fixed $n$, can be further factorized using the eigen basis of the transition matrix, one readily finds that the coefficients $c_f$ are positive. For example in the MF approach this is realized in terms of matrix elements $\langle \obs_\alpha|n \rangle$ with $|n\rangle$ the eigenvectors of the Hamiltonian (the MC case is discussed in Appendix~\ref{app:gamma}).
Note that in principle one could also examine the complete autocorrelation function along both MC and MF directions. 
In Appendix~\ref{app:MC_MF} we present a numerical result and discuss a few possible future directions, but its study, and more specifically the formulation of a multi-dimensional bounding strategy, goes beyond the scope of this paper.

\subsection{Improved bounds}

When the knowledge of the largest modes $\tau_0, \tau_1, \dots$ together with the corresponding coefficients $c^n_{\alpha\beta}$ is available, e.g. from variational methods, tighter bounds are obtained by subtracting the leading exponentials from $\Gamma_{\alpha \beta}(t)$. This method~\cite{Bruno:2019nzm} has been successfully exploited in the calculation of the HVP contribution to the muon anomaly~\cite{RBC:2024fic, Djukanovic:2024cmq, Bazavov:2024eou} and here it plays a similar role.

In lattice field theory calculations, one typically considers several correlators from different interpolating operators with the same quantum numbers. This condition guarantees that the spectrum is unique and that they can be treated simultaneously, either by a simpler global fitting procedure or by solving the corresponding Generalized Eigenvalue Problem (GEVP) (see Refs.~\cite{Luscher:1990ck, Blossier:2009kd}), provided that a (hermitian) square matrix of correlators can be constructed. This typically leads to a robust knowledge of the spectrum.

Here we notice that the autocorrelation function $\Gamma_{\alpha\beta}(t)$ in Eq.~\eqref{eq:gamma_exp} shares the same properties. 
Therefore, the bounding method introduced above can be straightforwardly improved by finding the eigenvalues $\lambda_n(t,t_0)$ and eigenvectors $v_{\beta n}(t,t_0)$ from the solution of the GEVP
\begin{equation}\label{eq:GEVP_gamma}
    \sum_\beta \Gamma_{\alpha\beta}(t) \, v_{\beta n}(t,t_0) = \lambda_n(t,t_0) \sum_\beta \Gamma_{\alpha\beta}(t_0) \, v_{\beta n}(t,t_0)\,, \quad t>t_0 \,.
\end{equation}
In the equation above $t$ refers to either physical time or markovian time, depending on the analysis that one is performing. 
Notice that the extracted effective overlaps (see Appendix~\ref{app:GEVP_mel_en}) are physical matrix elements in the former case, whereas in the latter they are interpreted as matrix elements in the sense of Eq.~\eqref{eq:gamma_spec_dec}. 
By constraining the tail of the autocorrelation function, the dominant terms at large times in Eq.~\eqref{eq:gamma_exp} can be determined systematically and consequently improve the bounding method, more specifically its upper bound, by saturating it at shorter windows. 

Some care is clearly needed here. As mentioned above a sufficient separation among the modes is the first important ingredient for the success of the GEVP. For  MF analysis the ideal setup is a lattice with spatial volumes of typical sizes, i.e. $m_\pi L \approx 4$, but very long time extent. Going to larger spatial extents makes the spectrum denser and compromises the applicability of the GEVP. For MC analysis less is known on the separation of the modes, which we would expect to decrease for systems admitting a larger configuration space. Therefore, in the presence of long Markov chains, when the autocorrelation function is known to sufficient accuracy, the GEVP may be useful mostly in providing an estimate of the slowest mode (which otherwise should be taken as the largest integrated autocorrelation time observed). 
Below we use it in a numerical demonstration precisely for this purpose, and in general we expect the use of variational methods in MC analysis mostly as additional help to solidify the knowledge of the slowest mode to be used in the upper bound, rather than a real improvement on the bounding method.

Notice also that when several shorter replicas, or several insufficiently long master-field configurations are available, variational methods should be used with care as well, since the autocorrelation function itself may suffer from systematic biases: for example, in the physically more intuitive picture offered by the one-dimenionsal MF analysis, $\Gamma_{\alpha\beta}(t)$ may be affected by thermal effects like the other correlation functions of the theory.

\section{Numerical tests}\label{sec:Results}

In this Section we perform a few numerical tests of the ideas introduced above. First we analyze synthetic autocorrelated Monte Carlo data and then we turn the discussion to real quantum field theories by focusing on a few observables of the $\CPN$ model, a two-dimensional theory sharing several similarities with QCD such as asymptotic freedom and confinement~\cite{DADDA197863,Campostrini:1992ar}, and the $\mathrm{SU}(3)$ pure-gauge theory. The latter, is worth considering not only for being a confining 4D gauge theory but especially for testing our ideas for the MF analysis in the presence of smearing, a common practice in modern calculations, for example for scale setting. Simulations have been performed using the \texttt{Grid}~\cite{GRID,Boyle:2016lf} and \texttt{gpt}~\cite{GPT} software libraries.\footnote{For the $\CPN$ model a second C++ code has been used for cross-checks.}

\begin{figure}[ht]
	\centering
    \includegraphics[width=.48\textwidth,keepaspectratio]{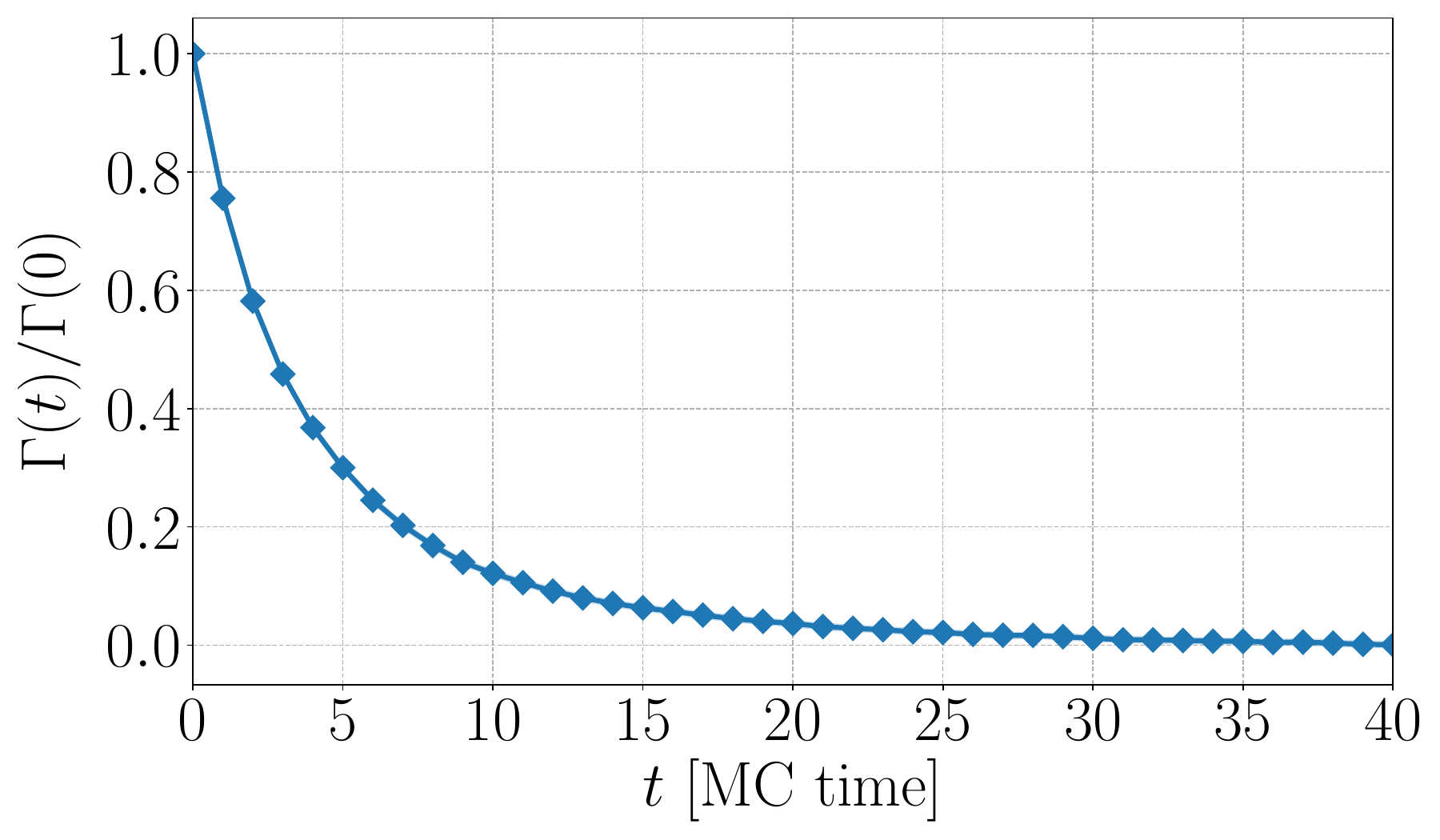}
    \includegraphics[width=.48\textwidth,keepaspectratio]{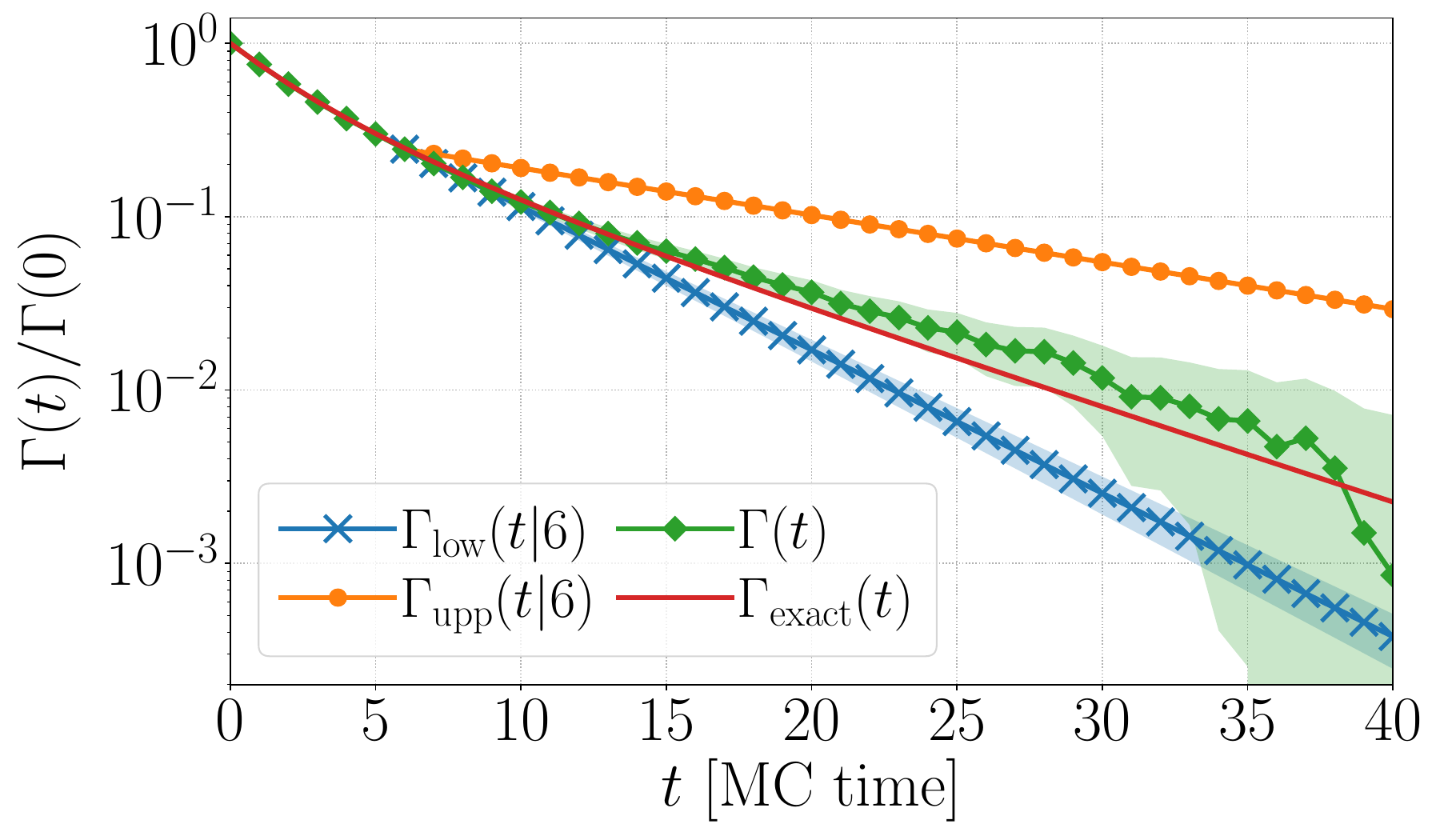}
	\caption{Numerical example with synthetic data containing three modes. \textit{Left:} the normalized autocorrelation function, also denoted with $\rho(t)$ in the literature. \textit{Right:} calculation of the upper and lower bounds of $\Gamma(t)$ for $W=6$ and comparison with both the estimated and exact analytic autocorrelation functions. For the upper bound the value $\widetilde{\tau}_0 = 16$ has been used.}
    \label{fig:Bounding_MC_synthetic}
\end{figure}

\begin{figure}[ht]
    \includegraphics[width=.48\textwidth,keepaspectratio]{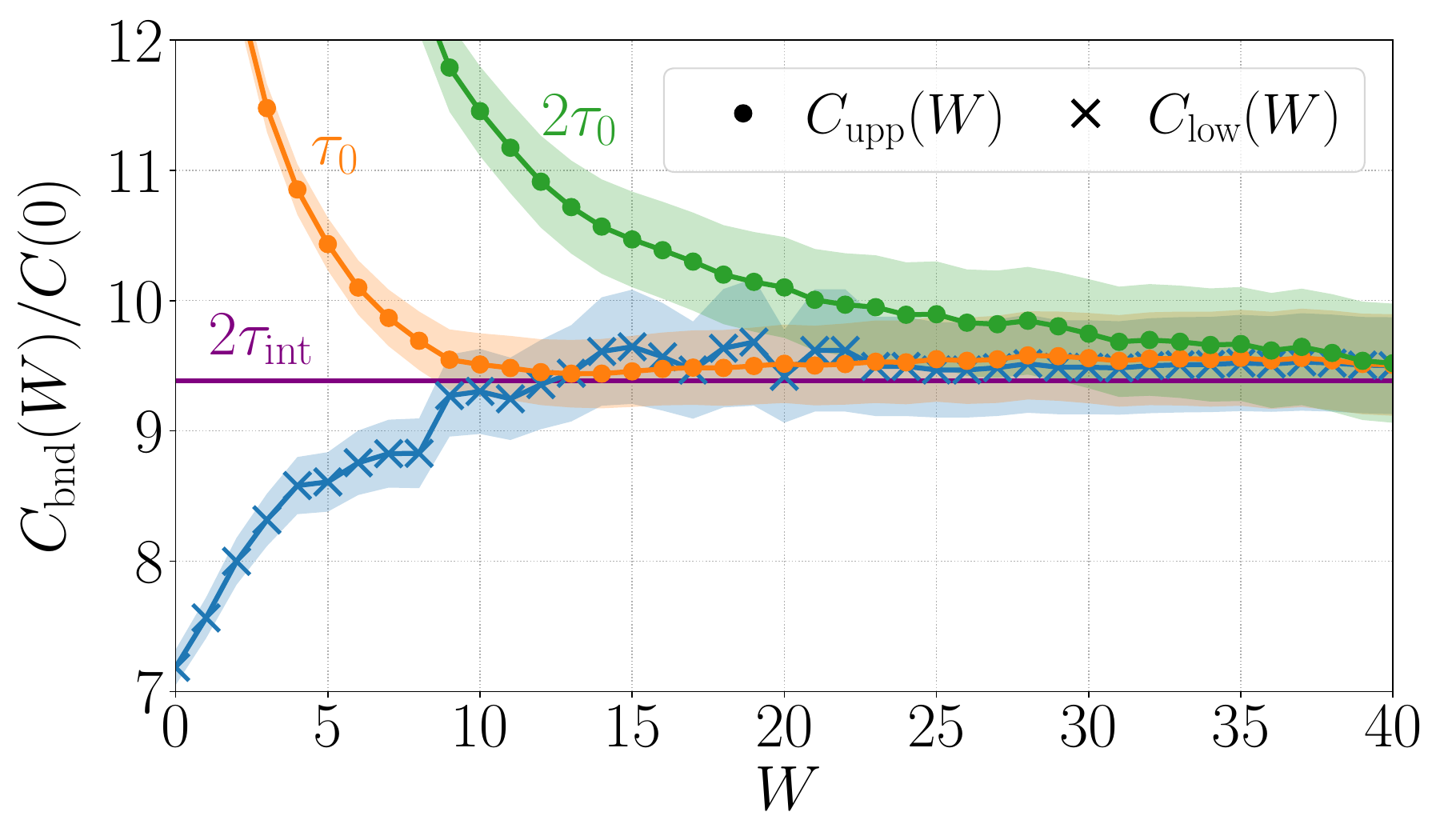}
    \includegraphics[width=.48\textwidth,keepaspectratio]{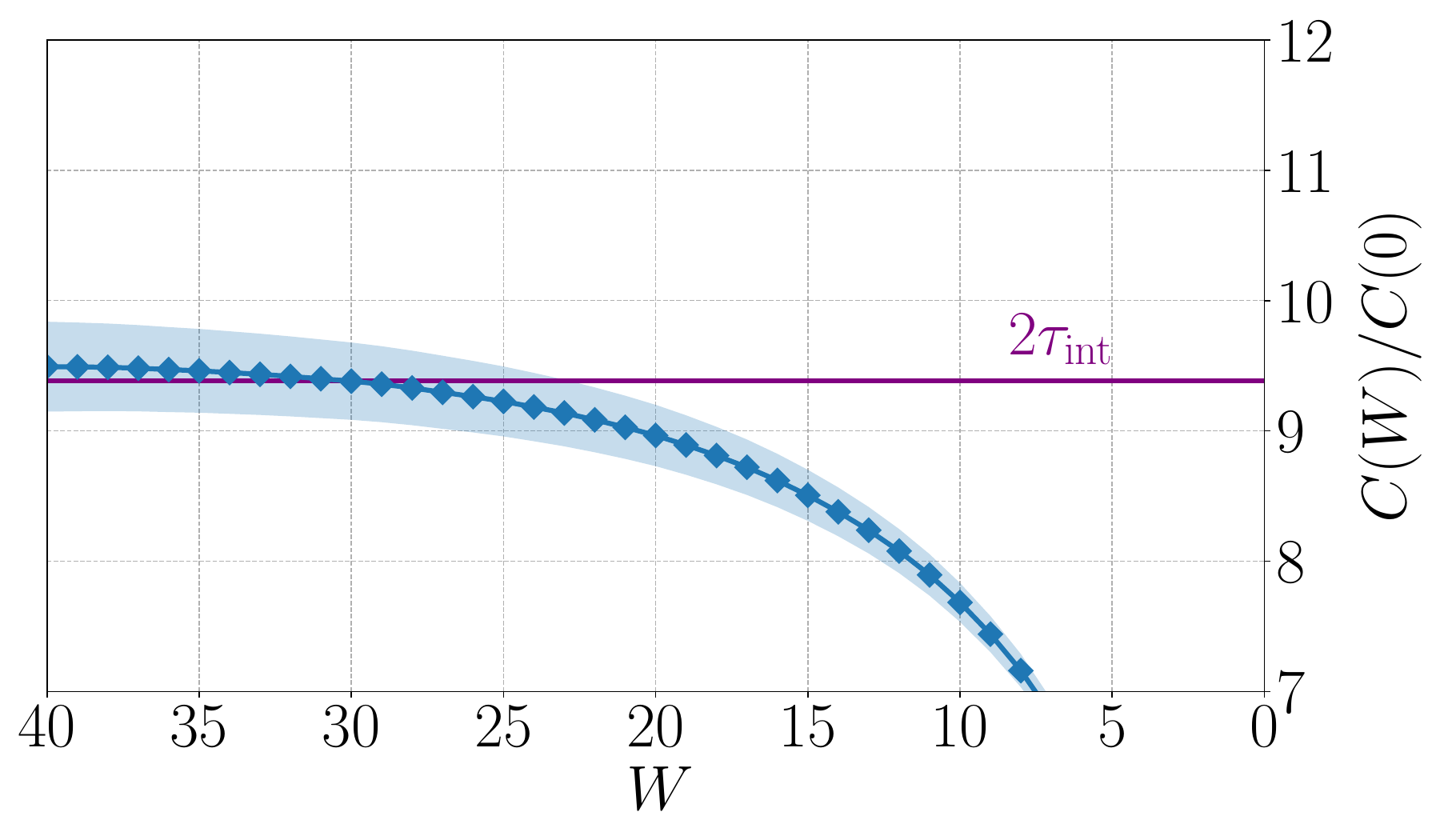}
	\caption{Application of the bounding method to synthetic autocorrelated data; the shaded bands represent the statistical errors of the employed estimators. Plateaus in both plots tend to the expected value $2 \tau_\mathrm{int}$. \textit{Left:} upper and lower bounds of the integrated autocorrelation function. To test the stability of the procedure we used $\widetilde \tau_0=8,16$ in the upper bound. For the former a good saturation of the bounds is achieved at $W \approx 10$, where $\Gamma(t)$ is still visibly decaying, implying a single mode dominance from this point on. For compactness, we indicate both upper and lower bounds with $C_\mathrm{bnd}(W)$ on the $y$-axis label. \textit{Right:} the integrated autocorrelation function as a function of the summation window.}
    \label{fig:Bounding_MC_synthetic_2}
\end{figure}

By using the procedure outlined in Ref.~\cite{Wolff:2003sm} and implemented in the \texttt{pyobs} software library~\cite{mattia_bruno_2024_14170562}, we have generated a mock Markov chain (with $10^5$ configurations) to perform the analysis on a single scalar observable $\obs$ (for simplicity we drop the lower indices $\alpha,\beta$).
For the underlying synthetic autocorrelation function we adopted
\begin{equation}
    \Gamma_\mathrm{exact}(t) = \frac13 \sum_{n=0,1,2} e^{-|t|/\tau_n}\,, \quad \tau_0 = 8 \,, \tau_1 = 4 \,,\tau_2 = 2 \,.
    \label{eq:Gamma_ex}
\end{equation}
According to Eq.~\eqref{eq:tau_int} its analytic $\tau_\mathrm{int}$ is therefore equal to $\approx 4.691$.
The plots in Fig.~\ref{fig:Bounding_MC_synthetic} report the application of the bounding method to the synthetic autocorrelation function estimated according to Eq.~\eqref{eq:estimator_G}. Its error is depicted by shaded bands and is calculated numerically as described in Section~\ref{sec:Bounding}.
As expected, given the large statistics, both the autocorrelation function and its area, reported in the left and right panels of Figs.~\ref{fig:Bounding_MC_synthetic} and \ref{fig:Bounding_MC_synthetic_2} respectively, are smooth functions of $W$ and well under control, a fact that makes the bounding windowing procedure clear and effective. 
In the right panel of Fig.~\ref{fig:Bounding_MC_synthetic} we see how the two bounds (in logarithmic scale) called for simplicity 
\begin{equation}
    \Gamma_\mathrm{upp}(t | W) \equiv \Gamma_\mathrm{bnd}(t | W , \widetilde \tau_0) \,, \quad \text{and} \quad \Gamma_\mathrm{low}(t | W) \equiv \Gamma_\mathrm{bnd}(t | W , \tau_\mathrm{eff}^W) \,,
\end{equation}
with $\widetilde \tau_0 \geq \tau_0$, have the expected behavior relative to the calculated autocorrelation function. 
Analogously we define the bounds on its integral as 
\begin{equation}
    C_\mathrm{upp}(W) \equiv \sum_{t=-\infty}^\infty \Gamma_\mathrm{upp}(t | W) \,, \quad \text{and} \quad C_\mathrm{low}(W) \equiv \sum_{t=-\infty}^\infty \Gamma_\mathrm{low}(t | W) \,,
\end{equation}
and study them as a function of the summation window, showing their saturation for large values of $W$.

In the application of the criterion in Eq.~\eqref{eq:stopping_criterium} (with $M=2$) we performed a stability test by choosing the value of $\widetilde \tau_0$ equal to $\{ \tau_0,  2\tau_0, 3\tau_0\}$ (with $\tau_0=8$ from Eq.~\eqref{eq:Gamma_ex}) thus obtaining automatic windows equal to $\{ 13, 26, 38\}$ respectively. As one can see from the results shown in Fig.~\ref{fig:Bounding_MC_synthetic_2} larger values of this mode lead to larger summation windows, as expected. Instead using the exact slowest mode in the upper bound provides a short effective window, $W=13$, where the autocorrelation function is still visibly non-zero, as one can see from the left panel of Fig.~\ref{fig:Bounding_MC_synthetic}. As commented already earlier, the criterion simply tells us that the autocorrelation function is dominated by a single mode for $t>13$, an effect captured by the two bounds. Compared to Wolff's procedure which returns $W=41$ this method allows to stop the summation much earlier.

\subsection{The $\CPN$ model}\label{subsec:cpn}

To test the method in a more realistic example we consider now the $\CPN$ model~\cite{Witten:1978bc,DAdda:1978vbw,Rabinovici:1980dn,DiVecchia:1981eh,Berg:1981er}, and more specifically the $\mathbb{CP}^9$ case, whose action on a 2-dimensional periodic lattice $T \times L$ and sites $x \equiv (x_0,x_1)$ is given by~\cite{Campostrini:1992ar, Engel:2011re}
\begin{equation}
    S[z^\dagger \! ,z,\lambda] = -2 \beta M\sum_x \sum_{\mu=1}^2 \left\{ \Re \! \left[ z^\dag(x+\hat{\mu}) z(x) \lambda_\mu(x) \right] - 1 \right\} \, .
\end{equation}
Note that $z(x)$ are unit-norm vectors in $\mathbb{C}^{M}$, $\lambda_\mu(x)$ are $\mathrm{U}(1)$ gauge link variables and $\hat{\mu}$ is the unit vector pointing towards the direction $\mu$. To keep a consistent notation with the rest of the paper we set the lattice spacing $a=1$. 

From the gauge-invariant operator $P_{ij}(x) \equiv z_i(x) z_j^\ast(x)$ for $i,j=1,\ldots, M$ we calculate the two-point correlation function $\Tr{\langle P(x) P(y) \rangle}_\mathrm{c}$, where the subscript indicates that we are considering connected correlators and the trace is taken in the $\mathbb{C}^{M}$ space.
For later convenience, we introduce first the correlator projected to zero spatial momentum in time-momentum representation
\begin{equation}
    G_2(x_0) = \Tr \langle \widetilde P(x_0) \, \widetilde P(0) \rangle_\mathrm{c}\,, \quad \text{with} \quad
    \widetilde P(x_0) = \frac{1}{L} \sum_{x_1} P(x) \,,
    \label{eq:G2}
\end{equation}
and then the Fourier transformed one
\begin{equation}
    \widetilde G (l_0) \equiv L \sum_{x_0} G_2(x_0) \, e^{2\pi i x_0 l_0/ T} \,, 
\end{equation}
where $l_0 \in \{0, \dots T-1 \}$ and the normalization with $L$ is inserted such that $\widetilde G(l_0)$ agrees with Refs.~\cite{Campostrini:1992ar,Engel:2011re}.
In our simulations, together with the energy density $E = \frac{S[z^\dagger \! ,z,\lambda]}{M \beta V}$, with $V=LT$,  we measure the magnetic susceptibility $\chi_\mathrm{M} = \widetilde{G}(0)$, the second-moment correlation length $\xi_\mathrm{G}$~\cite{Campostrini:1992ar}
\begin{equation}\label{eq:xi_G}
    \xi^2_\mathrm{G} = \frac{1}{4 \sin^2 \! \left( \frac{\pi}{T} \right)} \left[ \frac{\widetilde{G}(0)}{\widetilde{G}(1)} - 1 \right] \, ,
\end{equation}
as well as the global topological charge $Q=\sum_{x} q(x)$, from the density~\cite{Berg:1981er}
\begin{equation}
    q(x) = \frac{1}{2 \pi} \Im \log \Big\{ \Tr \! \left[ P(x + \hat{\mu} + \hat{\nu}) P(x + \hat{\mu}) P(x) \right] 
    \Tr \! \left[ P(x + \hat{\nu}) P(x + \hat{\mu} + \hat{\nu}) P(x) \right] \Big\} \, , \quad \text{with} \quad \hat{\mu} \neq \hat{\nu} \, ,
\end{equation}
and its susceptibility $\chi_\mathrm T = \langle Q^2 \rangle / V$.  \\

\textit{Monte Carlo analysis} --- To analyze data similar to Lattice QCD simulations, we have used the HMC algorithm~\cite{Duane:1987de} following Ref.~\cite{Engel:2011re}. In this realization it obeys the necessary properties, listed in the previous section, for the validity of the bounding method.
We have simulated a $72 \times 72$ lattice with $\beta = 0.85$ and collected $36000$ measurements\footnote{A total of $36K$ molecular dynamics units (MDU) have been generated, each integrated through $97$ leap-frog steps. We measured on every trajectory.}, whose results are summarized in Table~\ref{tab:CPN_obs_tau}. The choice of parameters corresponds to lattices where autocorrelations are noteworthy, but contrary to Ref.~\cite{Engel:2011re} we have employed lower statistics to be in a situation more similar to state-of-the-art lattice simulations. Notice that the $\CPN$ model has been extensively studied in the literature both from the analytic point of view, using e.g. large $M$ techniques~\cite{Witten:1978bc,DAdda:1978vbw}, or by lattice simulations, predominantly as a laboratory for exploratory algorithms, see for instance Refs.~\cite{Vicari:1992jy, Rindlisbacher:2016cpj, Flynn:2015uma, Hasenbusch:2017unr}.

\begin{table}[ht]
	\centering
    \begin{tabular}{ccl}
        \toprule
        & exp. value & $\phantom{00}\tau_\mathrm{int}$ \\
        \midrule
        $E$      & $0.622337(39)$ & $\phantom{0}3.55(14)$ \\[.5ex]
        $\chi_\mathrm{M}$ & $46.89(20)$ & $\phantom{0}9.41(52)$ \\[.5ex]
        $\xi_\mathrm{G}$  & $6.389(53)$ & $\phantom{0}7.21(34)$ \\[.5ex]
        $10^5 \chi_\mathrm{T}$ & $45.5(3.3)$ & $\phantom{0}48.8(6.0)$ \\[.5ex]
        \bottomrule
    \end{tabular}
    \caption{Observables and corresponding integrated autocorrelation times obtained from the simulation of the $\mathbb{CP}^9$ model on a $72 \times 72$ lattice with $\beta = 0.85$. The values are in good agreement with Ref.~\cite{Engel:2011re}, while the errors differ as expected from the difference in statistics. Errors and $\tau_\mathrm{int}$'s have been calculated using the bounding method.}
    \label{tab:CPN_obs_tau}
\end{table}

\begin{figure}[ht]
	\centering
    \includegraphics[width=.45\textwidth,keepaspectratio]{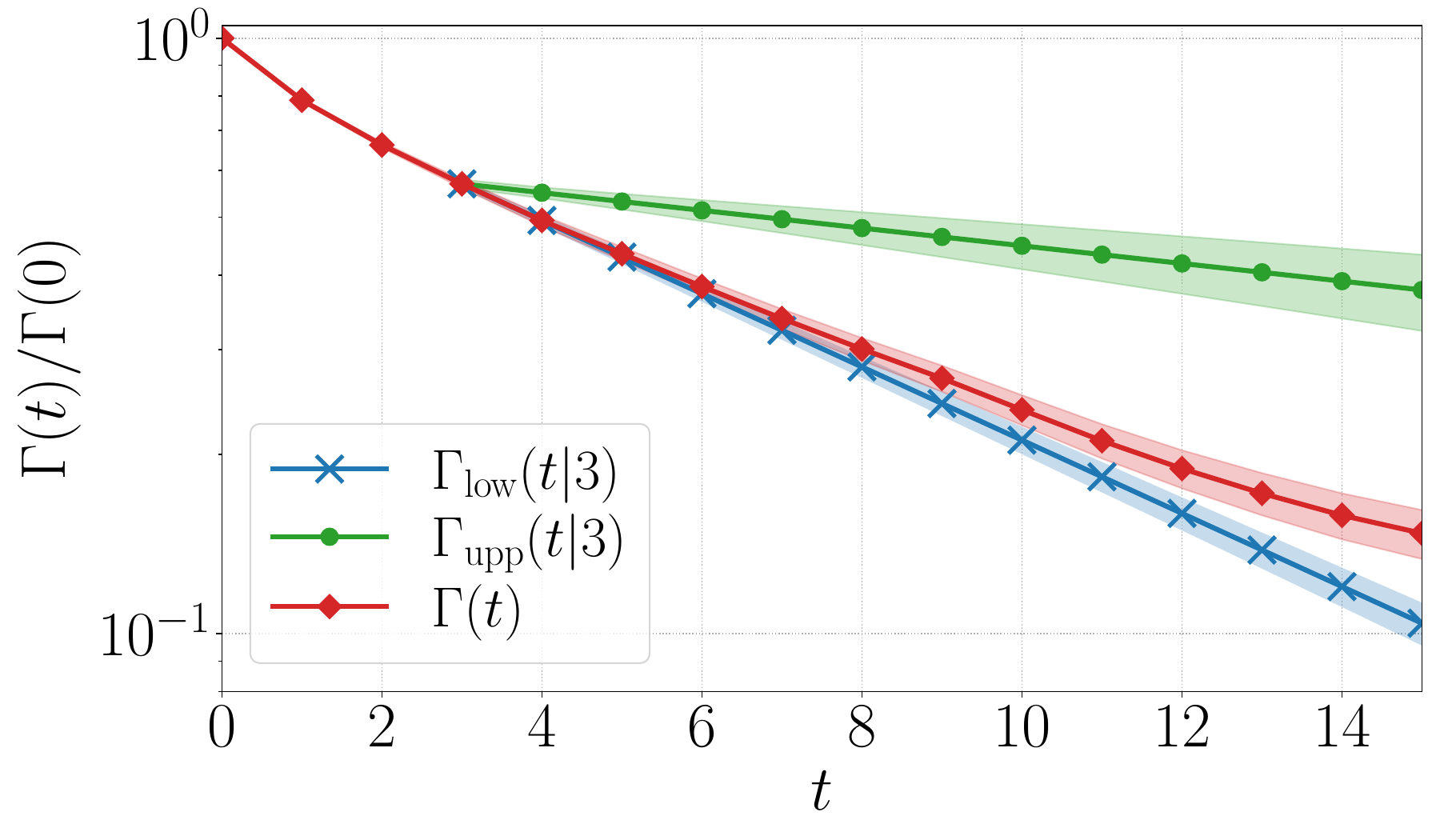}
    \includegraphics[width=.45\textwidth,keepaspectratio]{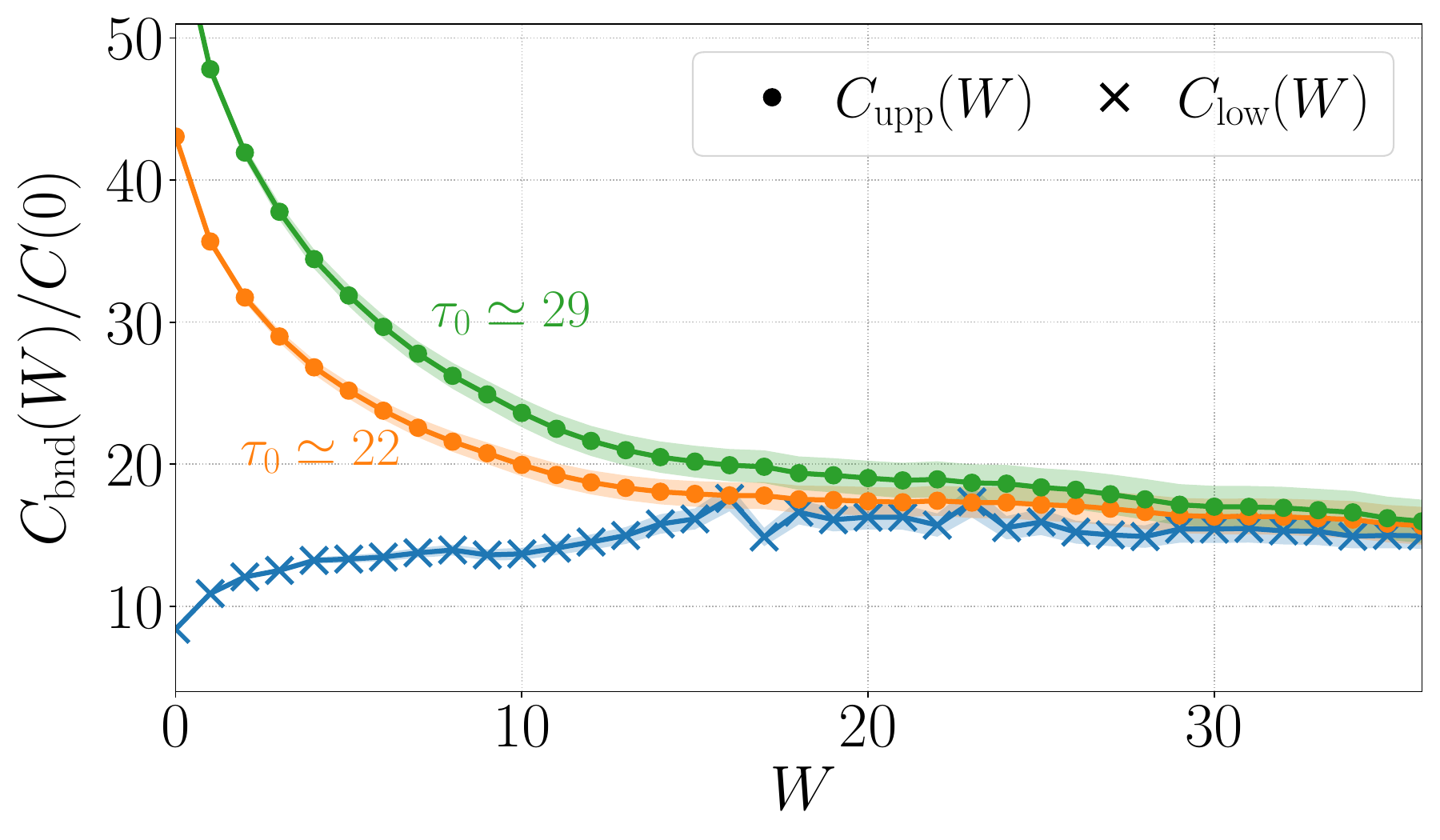}\\
    \includegraphics[width=.45\textwidth,keepaspectratio]{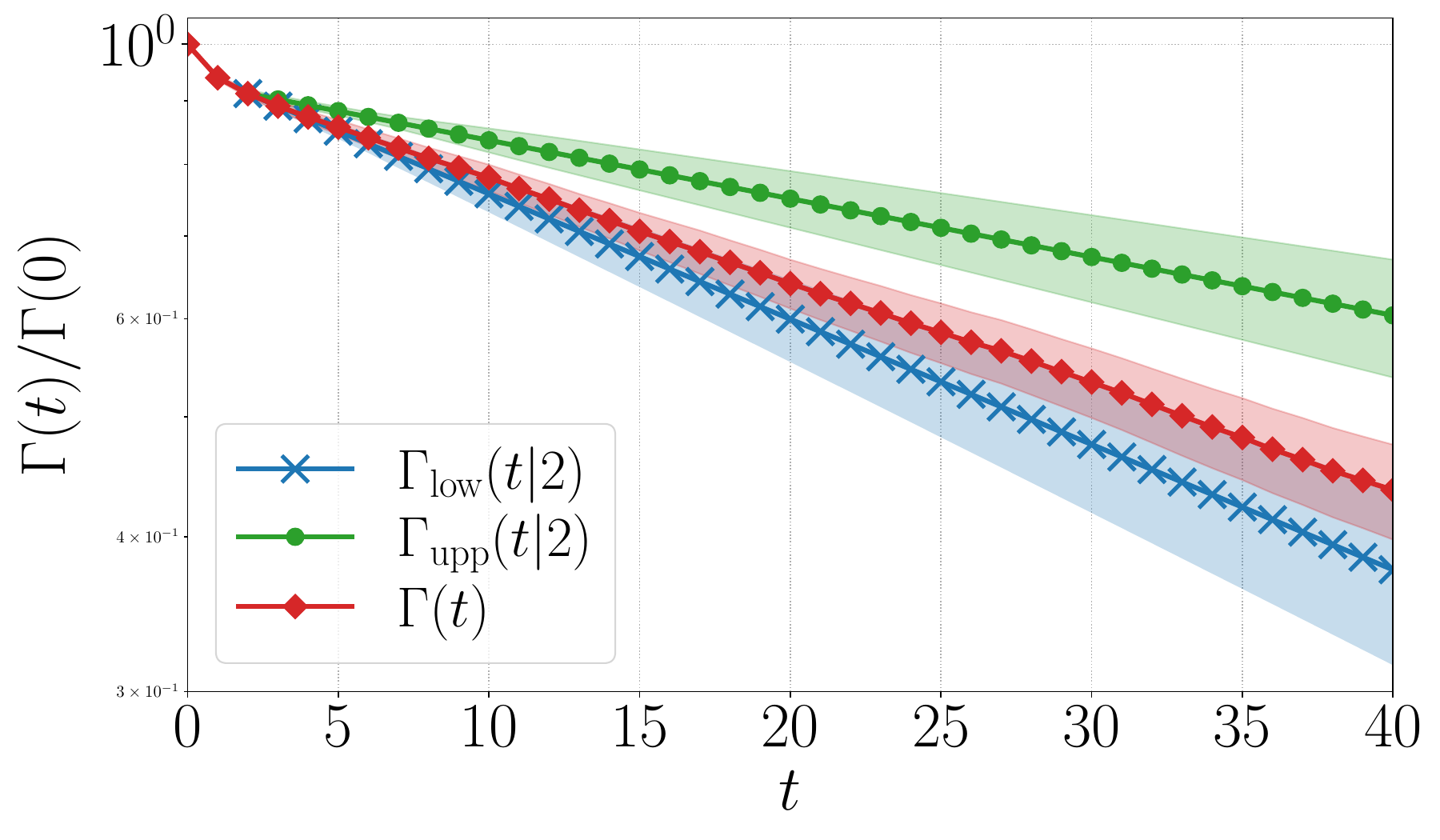}
    \includegraphics[width=.45\textwidth,keepaspectratio]{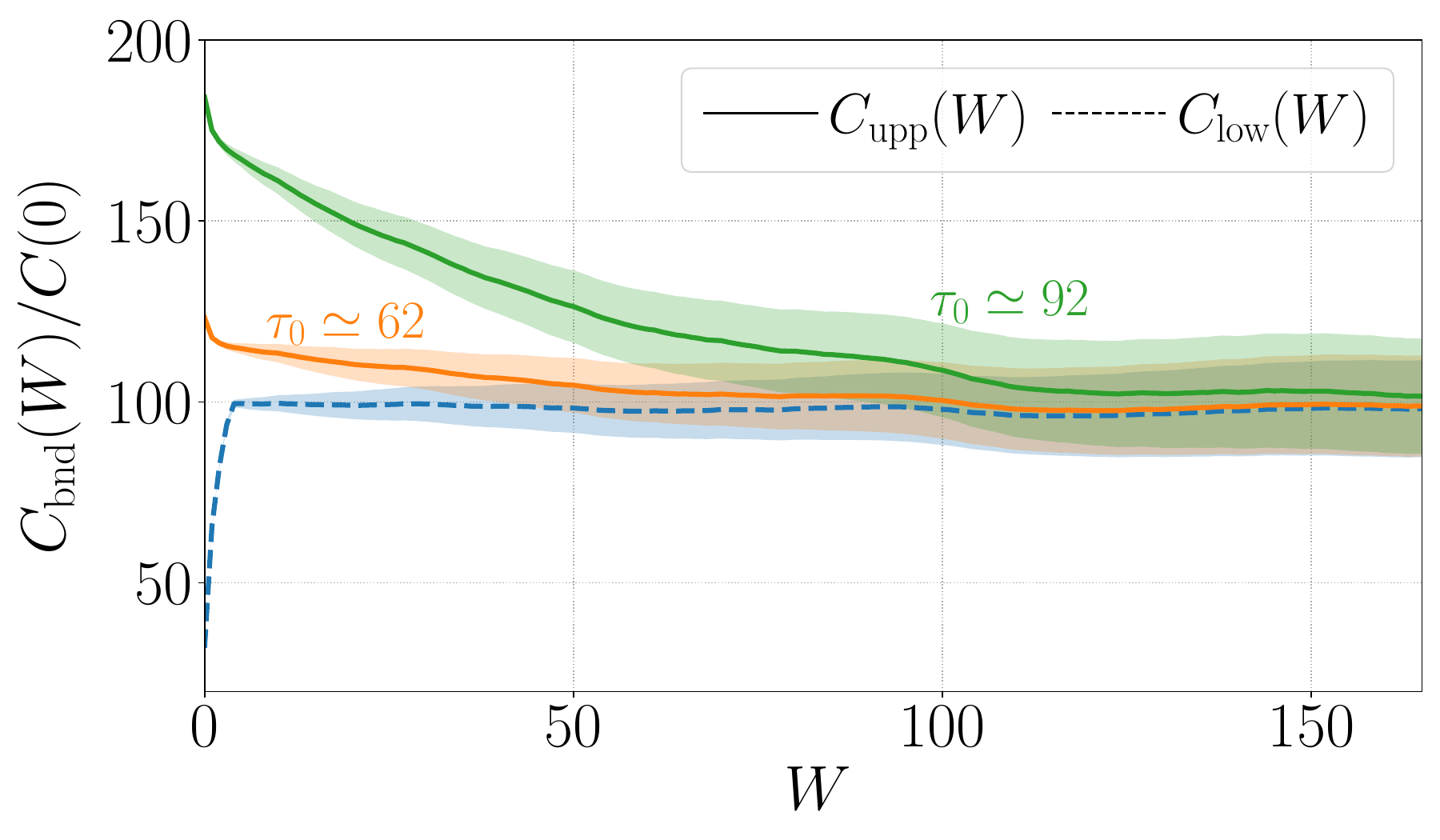}
    \caption{Bounding method applied to the Monte Carlo autocorrelation function of $\xi_\mathrm{G}$ and $\chi_\mathrm T$, in the top and bottom rows respectively. The shaded bands represent the statistical errors. \textit{Left}: bounds $\Gamma_\mathrm{upp}(t | W)$ and $\Gamma_\mathrm{low}(t | W)$ for the windows reported in the legends. The values of $\tau_0$ in $\Gamma_\mathrm{upp}$ correspond to the largest ones reported in the right panel, and are denoted by the same green color. \textit{Right}: integral of the bounds as a function of $W$. The orange and green curves correspond to $C_\mathrm{upp}(W)$ for the first and last iterations of the method described in the main text.}
    \label{fig:Bnd_MC_CPN}
\end{figure}

In Fig.~\ref{fig:Bnd_MC_CPN} we show a representative example of the application of the bounding method to the MC autocorrelation function of the correlation length $\xi_\mathrm{G}$ (top panels) and the magnetic susceptibility $\chi_\mathrm{T}$ (bottom panels). Its error, depicted by shaded bands in Fig.~\ref{fig:Bnd_MC_CPN} and used in the automatic procedure, is estimated according to Eq.~\eqref{eq:err_gamma}.
At first we focus on $\xi_\mathrm G$, neglecting the other measured observables (and their information on autocorrelations). We start by inspecting its lower bound. Using $W'=3$, $k=2$ and a sufficiently long summation window in Eq.~\eqref{eq:tau_hat}, we find $k \widehat \tau_0 \simeq 22$. By adopting it in the upper bound we obtain an optimal window of 16 and after iterating the procedure one more time we arrive at $k \widehat \tau_0 \simeq 29$ and $W=30$. At this point the iterative procedure stops, finding a satisfactory saturation of the autocorrelation function as reported in Fig.~\ref{fig:Bnd_MC_CPN}. Similarly we also test the simpler method based on the usage of $k \tau_\mathrm{int}$ in the upper bound and by choosing a multiplicative factor of 4 (and an estimate of $\tau_\mathrm{int}$ from the lower bound for the first iteration) we obtain a similar result, ending up with almost identical error estimates. As expected after a few iterations the window $W$ is stable in both approaches and the iterative procedure stops (for comparison Wolff's $\Gamma$-method returns $W = 56$).
Similarly we apply the same adaptive bounding methods to the observable with longest autocorrelation function observed, $\chi_\mathrm T$. 
The recursive procedure based on $\widehat \tau_0$ converges after a few iterations to a satisfactory window even with $k=1$, thanks to single-mode dominance which is evident from the flatness of the lower bound. From the right panels of Fig.~\ref{fig:Bnd_MC_CPN} we observe the expected behavior of the bounds and their saturation relative to their statistical errors. We remark again that both approaches depend on a few assumptions. In the better scenario (described below) where an estimate of $\tau_0$ is accessible the bounding method can be used straightforwardly, once.

We repeat the same operations on the other observables and show representative examples of the bounds in Appendix~\ref{app:obs_cpn}, while their estimated errors and integrated autocorrelation times are reported in Table~\ref{tab:CPN_obs_tau} and are in good agreement with Ref.~\cite{Engel:2011re}. Notably we find that the lower bound is often quite flat.

Finally, as a consistency check, we repeated the analysis on all observables by setting $\tau_0$ equal to the largest integrated autocorrelation time observed, i.e. of $\chi_\mathrm T$, and found very good stability in our procedure. Depending on the available data, one may decide to choose either the iterative procedure or this approach.\\

\begin{figure}[ht]
    \centering
    \includegraphics[width=.45\textwidth,keepaspectratio]{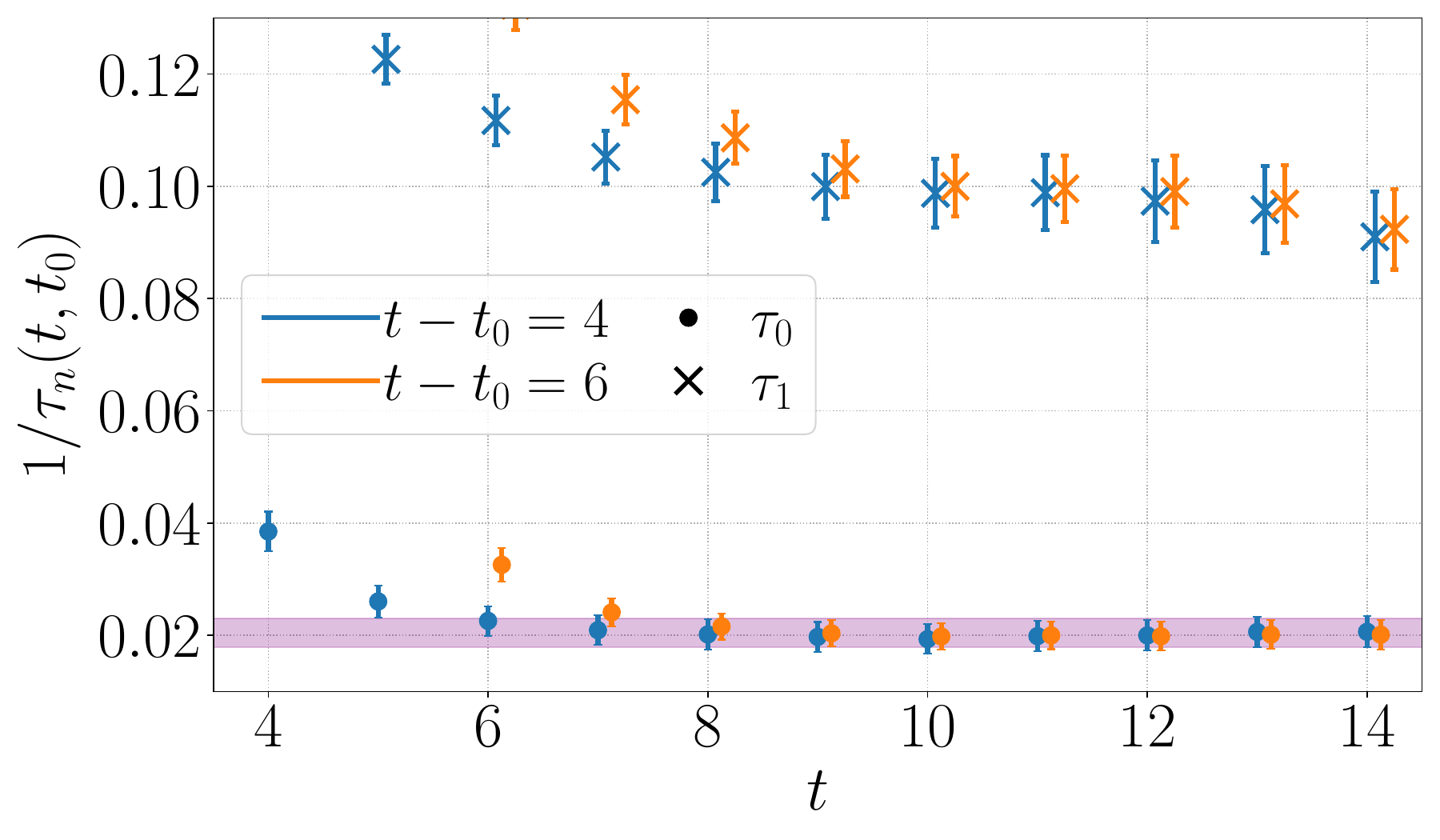}
    \includegraphics[width=.45\textwidth,keepaspectratio]{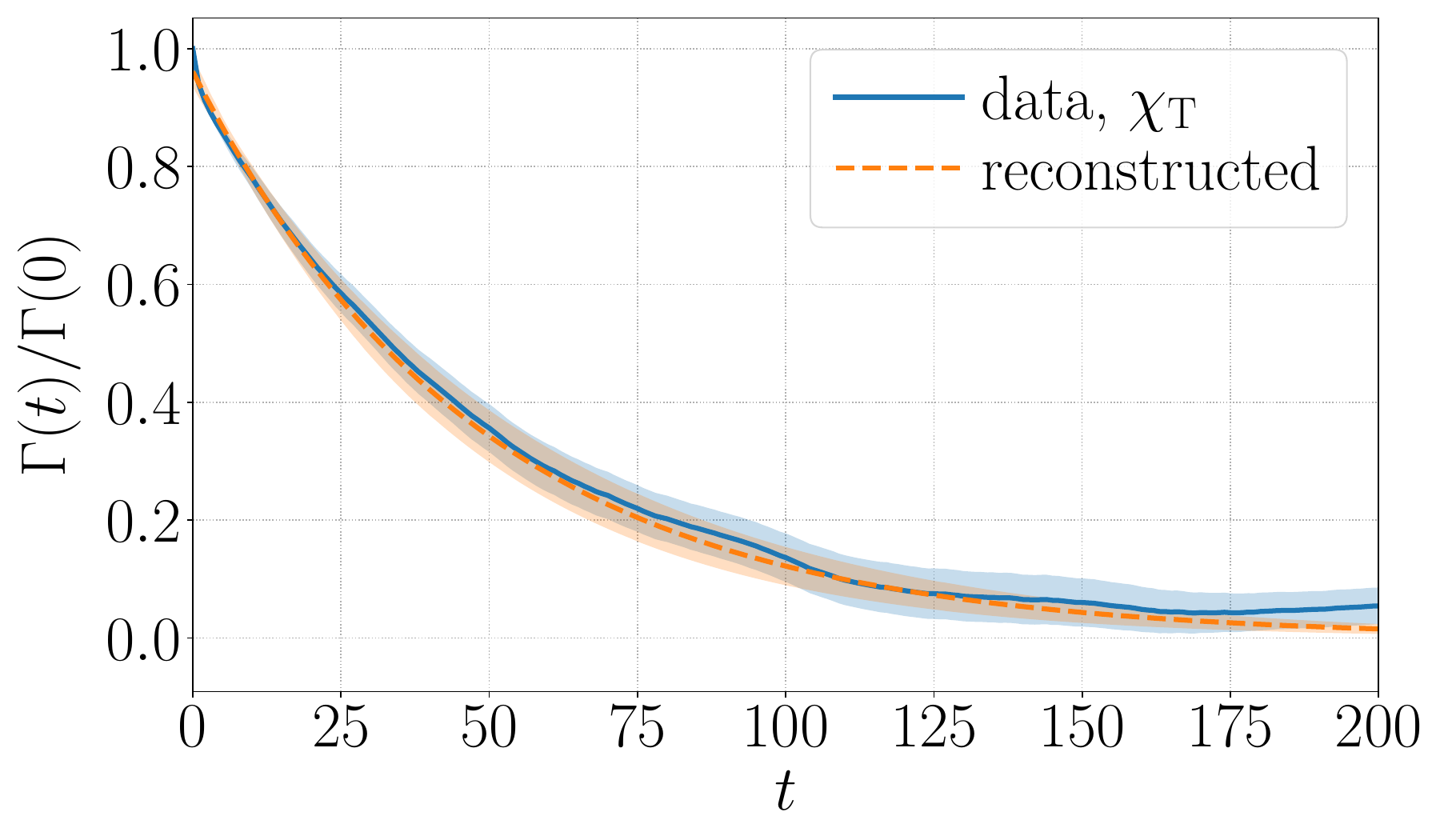}
    \caption{\textit{Left}: autocorrelation modes extracted from the projected $2 \times 2$ GEVP applied to the autocorrelation function $\Gamma_{\alpha\beta}(t)$ of 4 observables. The projection is taken at $t_\mathrm{P}=4$, see Appendix~\ref{app:GEVP_mel_en}. The purple horizontal band represents the (inverse) integrated autocorrelation time of $\chi_\mathrm{T}$, cf. Table~\ref{tab:CPN_obs_tau}. The orange points are slightly shifted along the $x$-axis for better readability. \textit{Right}: autocorrelation function of the topological susceptibility. The estimate obtained from data is compared to the reconstruction with the lowest exponential extracted from the pGEVP at $t=14$ for $t-t_0=4$ presented in the left panel.}
    \label{fig:GEVP_MC_CPN}
\end{figure}

\textit{Variational approach} --- To further solidify the error analysis presented above, and more specifically the value of $\tau_0$ used in the upper bound, we have also investigated the role of the GEVP. To this end we have calculated the correlator matrix $\Gamma_{\alpha\beta}(t)$ from the estimators of $E(t), \chi_\mathrm{T}(t), \chi_\mathrm{M}(t)$ and $\xi_\mathrm{G}(t)$ according to Eq.~\eqref{eq:estimator_G}. 
In typical lattice calculations of this type, guidance from our physical intuition is often very useful in designing a basis of orthogonal operators. When this is less clear, or when statistical errors are large, the generalized eigenvalue problem may become singular since $\Gamma_{\alpha\beta}(t)$ may not contain statistically independent information on all required states. To mitigate this problem we adopt a projection strategy (pGEVP), which is described in Appendix~\ref{app:GEVP_mel_en}, and apply it to the $4 \times 4$ GEVP, built from the four observables given in Table~\ref{tab:CPN_obs_tau}, to reduce it to a $2 \times 2$ pGEVP. Our results on the ``spectrum'', namely on the autocorrelation modes, are reported in the plots of Figs.~\ref{fig:GEVP_MC_CPN}. A comparison between the full and projected GEVP is reported in Appendix~\ref{app:GEVP_mel_en}, and excellent consistency is found between the two methods. 
Using two different approaches, where we fix either $t_0$ or $t-t_0$, we find a stable extraction of the lowest autocorrelation mode, which we find to be in good agreement with the integrated autocorrelation time of $\chi_\mathrm T$, underlying the dominance of this single mode in its autocorrelation function. 
This conclusion is further corroborated by the right panel of Fig.~\ref{fig:GEVP_MC_CPN} where we compare the autocorrelation function with the reconstruction obtained from the single mode $\tau_0$.
Contrary to $\chi_\mathrm T$, the other three observables couple strongly to the excited modes, which may however be affected by larger systematics and therefore we do not use to improve the bounding strategy. As remarked earlier, in the MC analysis it is advisable to use such a variational approach only to obtain an estimate of the longest mode to be used in the upper bound for the calculation of the error, as we do here.\\

\begin{figure}[ht]
    \centering
    \includegraphics[width=.45\textwidth,keepaspectratio]{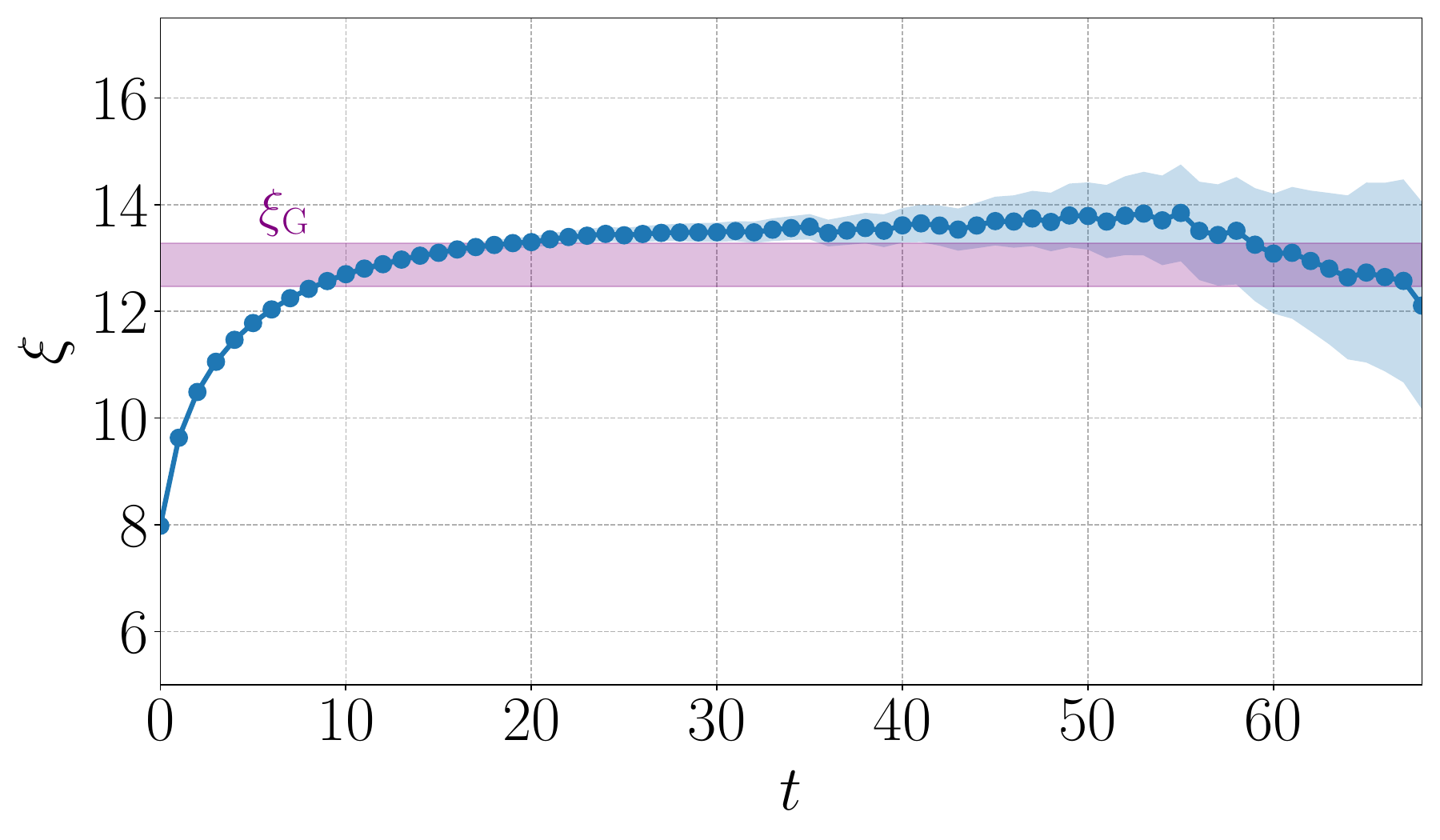}
    \includegraphics[width=.45\textwidth,keepaspectratio]{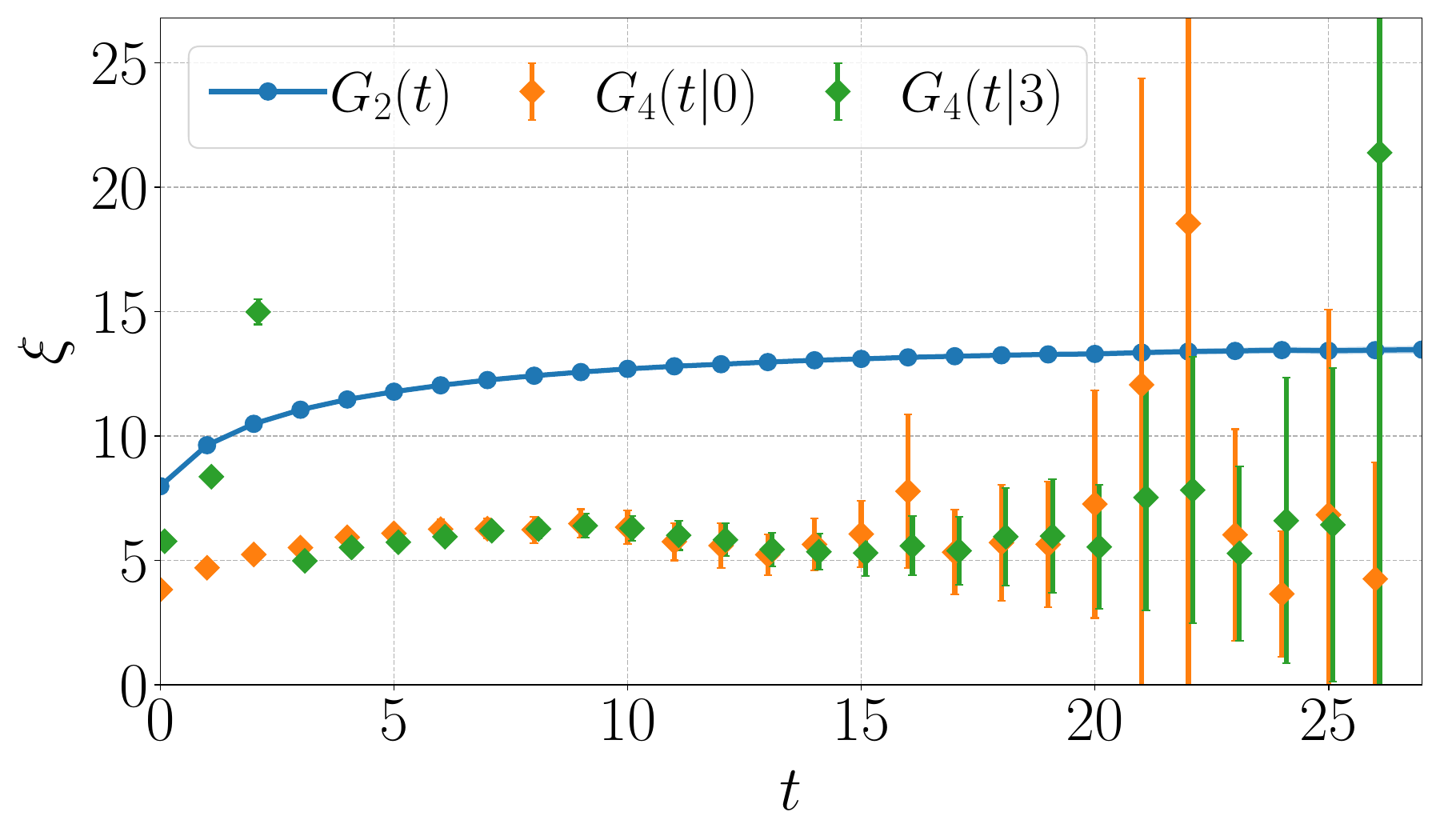}
    \caption{\textit{Left}: inverse mass gap $\xi$ extracted from the asymptotic fall-off of $G_2(x_0)$ (blue curve). The horizontal purple band represents the correlation length $\xi_\mathrm{G}$ defined according to Eq.~\eqref{eq:xi_G}. \textit{Right}: comparison of the inverse effective masses of $G_2(x_0)$ and $G_4(t|x_0)$ in the $x_0$ and $t$ coordinates respectively (points are slightly shifted for better readability). For the latter a spectral decomposition according to Eq.~\eqref{eq:gamma_exp} is achieved at times $t>x_0$ explaining the behavior of the first points for $G_4(t|3)$.}
    \label{fig:G2_G4}
\end{figure}

\textit{Master-field analysis} --- To test the bounding method in the context of the MF analysis we simulated a $240 \times 60$ lattice at $\beta = 0.95$ and studied the (auto)correlation function along the longer direction, $T$, using $240$ independent fields\footnote{We obtained such fields by skipping 100 MDU, namely four times the longest observed integrated autocorrelation time of $G_2(x_0)$. Each HMC trajectory has been integrated with $170$ leap-frog steps.}. At this value of $\beta$ the time extent of the lattice accommodates approximately 18 correlation lengths. The observable that we examine is the Euclidean two-point function introduced in Eq.~\eqref{eq:G2}. More specifically we consider the estimator
\begin{equation}
    \overline G_2(x_0) = \frac{1}{T} \sum_{t} \Tr \big[ \delta \widetilde P(x_0+t) \, \delta \widetilde P(t) \big] \,, 
\end{equation}
where the fluctuations $\delta \widetilde P(x_0)$ are defined in accordance with Eq.~\eqref{eq:fluctuations}.
Notice that above, the index $t$ spans different source locations, while $x_0$ labels the distance between the operators (i.e different observables and therefore it corresponds to the index $\alpha$ used in Section~\ref{sec:Bounding}). 
In the limit $T\to\infty$, the estimator $\overline G_2(x_0)$ tends to the connected zero-momentum correlator  given in Eq.~\eqref{eq:G2}.
The mass gap of the theory is extracted from its long-distance behavior and its inverse is indicated by $\xi$:
\begin{equation}
    \frac{1}{\xi} = \lim_{x_0 \to \infty} \log \! \left\{ \frac{G_2(x_0)}{G_2(x_0+1)} \right\} \,.
\end{equation}
The ratio $\xi_\mathrm{G}/\xi$ has a well-defined continuum limit and is predicted to deviate from 1 as $M$ increases~\cite{Campostrini:1991kv,Campostrini:1992ar}.
In the left panel of Fig.~\ref{fig:G2_G4} we show the extraction of the inverse mass gap, for which we find $\xi= 13.57(30)$ from a plateau average performed from $t=25$, to be compared with the measured value of $\xi_\mathrm{G} = 12.87(41)$\footnote{Relatively to Ref.~\cite{Flynn:2015uma} we find slightly different values, due to the different lattice geometries. By matching the lattice size, $120 \times 120$, at this value of the coupling, $\beta = 0.95$, we have reproduced the same values from a dedicated simulation, thereby confirming that these quantities suffer from relatively large finite-size effects.}.

The error of $\overline{G}_2(x_0)$
is calculated from the integral of the following connected correlation function
\begin{equation}\label{eq:G4}
    G_4(t|x_0) = \langle \Tr \big[ \delta \widetilde P(x_0+t) \, \delta \widetilde P(t) \big] \, \Tr \big[ \delta \widetilde P(x_0) \, \delta \widetilde P(0) \big] \rangle_\mathrm{c} \,.
\end{equation}
In terms of the notation used in Section~\ref{sec:Bounding}, $G_4(t|x_0)$ corresponds to $\Gamma_{\alpha\alpha}(t)$, with $\alpha$ spanning the values of $x_0$, i.e. different observables.
When $t>x_0$ then $G_4(t|x_0)$ becomes a typical two-point correlator with spectral representation as in Eq.~\eqref{eq:gamma_exp} and positive-definite matrix elements. The corresponding non-local operators $\Tr \big[ \delta \widetilde P(x_0) \, \delta \widetilde P(0) \big]$ are gauge and group invariant, but, compared to $\delta \widetilde P(x_0)$ entering in $G_2(x_0)$, transform under a different representation (singlet) of the global symmetry group $\mathrm{SU}(M)$. As a consequence a different spectrum with heavier asymptotic states is expected, as discussed in Appendix~\ref{app:cpn}, and we observe this in the right panel of Fig.~\ref{fig:G2_G4} by looking at the inverse effective masses of both $G_2(x_0)$ and $G_4(t|x_0)$ in the $x_0$ and $t$ coordinates respectively.
We conclude that using the correlation length $\xi$ in the bounding method is a safe and rather conservative approach here, and a similar situation may be present in QCD as well for those cases where two-pion states contribute to the MF autocorrelation function at long distances, and one chooses instead the pion mass in the upper bound.

\begin{figure}[t]
    \centering
    \includegraphics[width=.45\textwidth,keepaspectratio]{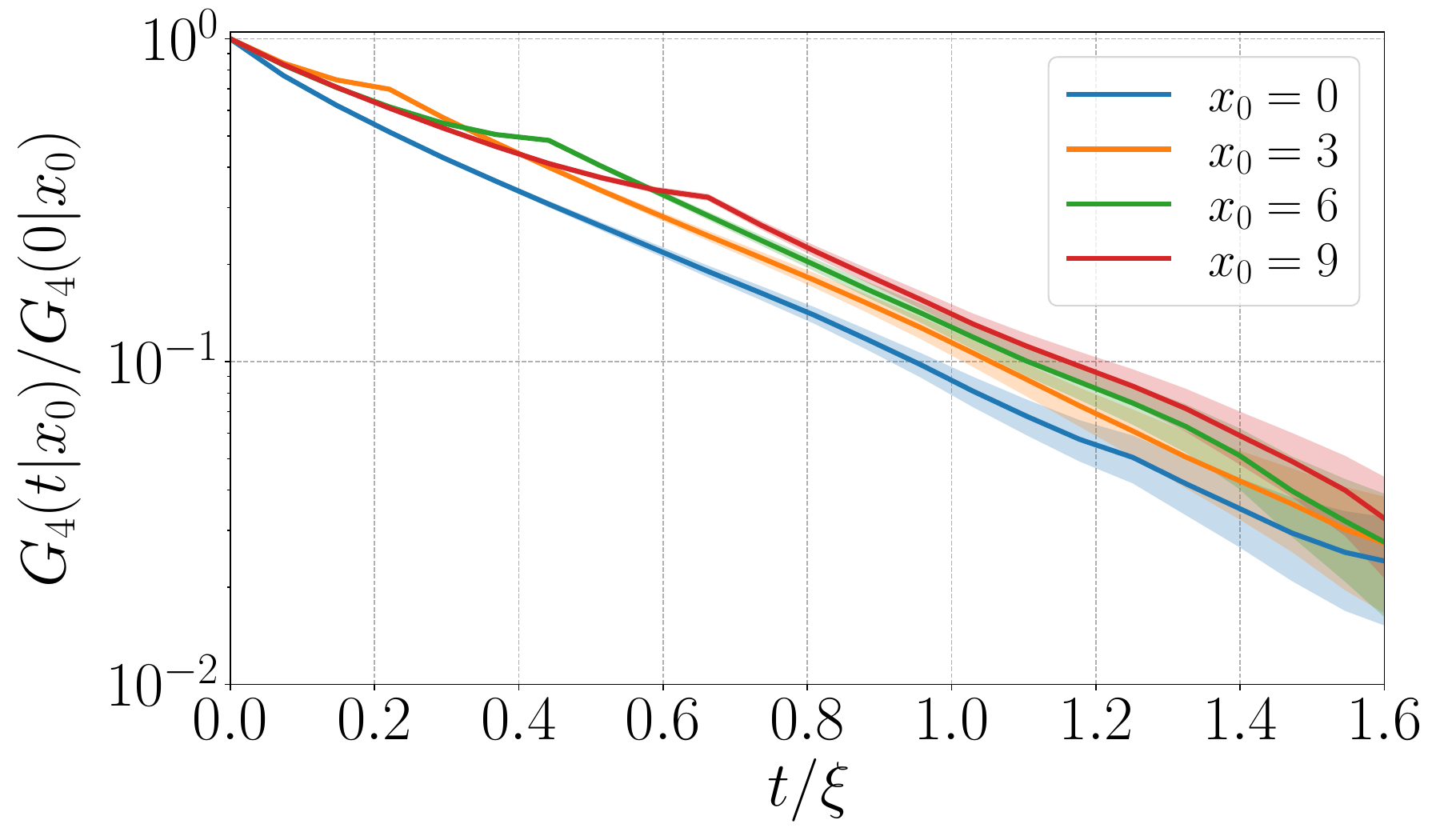}
    \includegraphics[width=.425\textwidth,keepaspectratio]{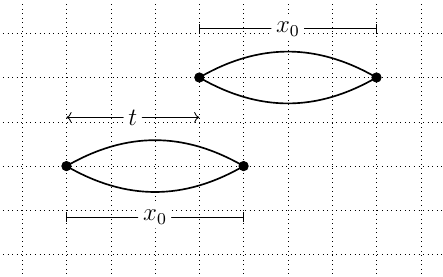}
    \caption{\textit{Left}: $G_4(t|x_0)$ for several values of $x_0$. When $t > x_0$ the spectral decomposition in Eq.~\eqref{eq:gamma_exp} is restored. \textit{Right}: Sketch of the two non-local operators contributing to the four-point function $G_4(t|x_0)$. The two blobs represent the traces in Eq.~\eqref{eq:G4} with $x_0$ the separation among the two operators $\delta \widetilde P$, within a single trace, and $t$ their relative distance.}
    \label{fig:lattice_G2_G4}
\end{figure}

\begin{figure}[ht]
    \centering
    \includegraphics[width=.45\textwidth,keepaspectratio]{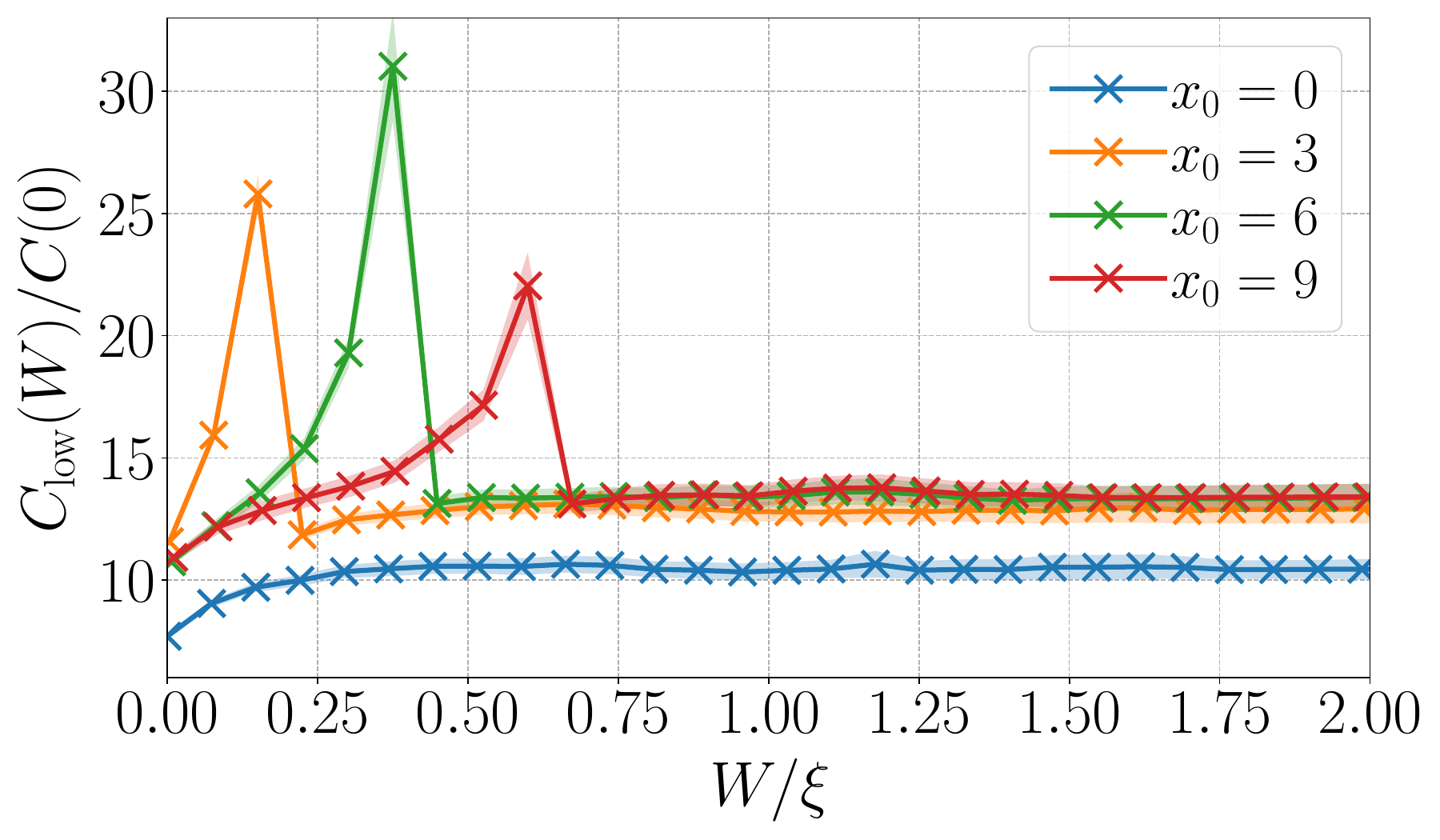}
    \includegraphics[width=.45\textwidth,keepaspectratio]{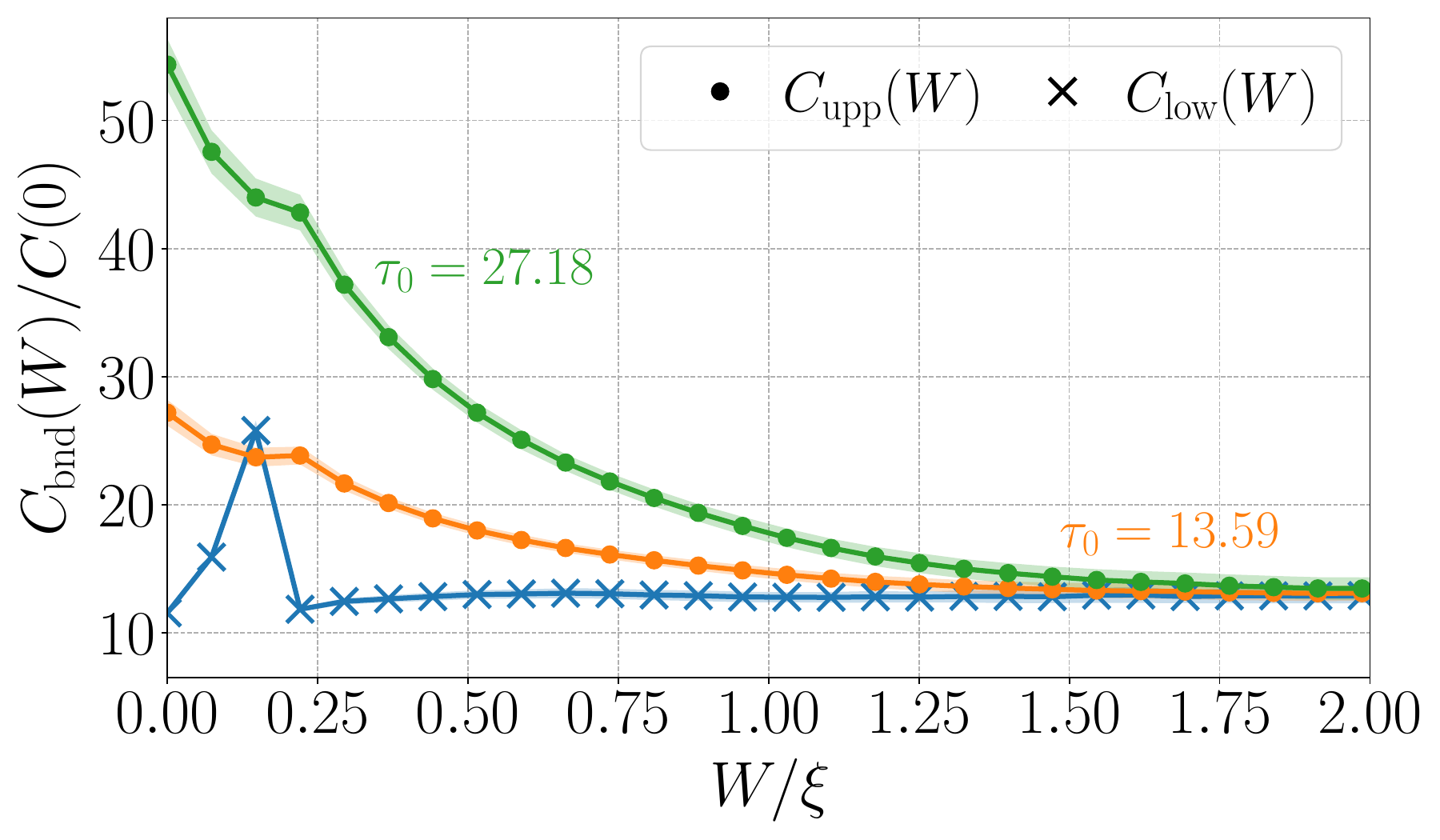}
    \caption{\textit{Left}: lower bound of the integrated autocorrelation function $G_4(t|x_0)$ for several values of $x_0$. \textit{Right}: bounding method applied to $G_4(t|3)$ with the upper bound calculated using $\tau_0 = \xi$ (orange curve) and $\tau_0 = 2 \xi$ (green curve).}
    \label{fig:G4_bounds}
\end{figure}

When $t \leq x_0$ the two non-local traced operators in $G_4(t|x_0)$ are still overlapping as represented with an illustration in the right panel of Fig.~\ref{fig:lattice_G2_G4}. In turn this translates into two different exponential behaviors of $G_4(t|x_0)$ for values of $t$ larger or smaller than $x_0$ and consequently two different regimes for the lower bound, as reported in the left panel of Fig.~\ref{fig:G4_bounds}. 
In fact, its steep rise for $t \lesssim x_0$ originates from the behavior of the effective mass below the discontinuity at $t=x_0$. We show it as a warning and in numerical applications we explicitly sum the autocorrelation function up to a minimal $t$ and apply the bounding method for larger values, where Eq.~\eqref{eq:gamma_exp} is valid.
In the right panel of Fig.~\ref{fig:G4_bounds} we compare the upper and lower bounds of $G_4(t|3)$ with the choices $\tau_0 = \xi$ and $2 \xi$ to demonstrate more explicitly such a failure and the importance of an explicit summation for $t \leq x_0$.
Once this initial part (up to $t=x_0$) is discarded we find the bounding method extremely stable and within our setup we can safely estimate the error of $G_2(x_0)$ up to $x_0 \simeq 40$, thanks to the long time extent of the lattice.
The error of the error used in the stopping criterion is calculated from the Monte Carlo direction, but alternatively one could use the analytic approximations mentioned in Section~\ref{sec:Bounding}. Finally, we remark that for local observables, such as $G_2(x=0)$, the bounding method works as expected without these complications.

\def\tf{t_\mathrm{f}}
\def\mG{m_G}

\subsection{Pure gauge $\mathrm{SU}(3)$ theory}

For the pure gauge theory we used the standard Wilson discretization on a $T \times L^3$ lattice \cite{Wilson:1974sk}
\begin{equation}
    S_\mathrm{W}[U] = \beta \sum_{x} \sum_{\mu > \nu} \left( 1 - \frac{1}{3} \Re \Tr U_{\mu \nu}(x) \right) \, , \quad \beta \equiv \frac{6}{g_0^2} \, ,
\end{equation}
where $g_0$ is the bare coupling, $U_{\mu \nu}(x) = U_\mu(x) U_\nu(x + \hat{\mu}) U_\mu^\dag(x+\hat{\nu}) U_\nu^\dag(x)$ is the standard plaquette  and $U_\mu(x)$ are the $\mathrm{SU}(3)$ gauge links connecting the site $x$ to $x+\hat{\mu}$ (the trace is over the color index).

A standard practice in modern simulations is to smear gauge links by solving a gradient flow equation along a fictitious flow time $\tf$~\cite{Luscher:2010iy}. Note that to avoid confusion with the variable $t$ used throughout the manuscript, we adopt $\tf$ which is different from the typical literature on the subject. 
In our calculation we employ the gradient of the Wilson action~\cite{Luscher:2009eq, Luscher:2010iy} and calculate the smeared link variables $V_{\mu}(x|\tf)$ for $\tf>0$ (such that $V_\mu(x|0) = U_\mu(x)$) from numerical solutions of the differential equation
\begin{equation}
    \dv{\tf} V_{\mu}(x|\tf) = -g_0^2 \left\{ \partial_{x,\mu} S_\mathrm{W}[V] \right\} V_{\mu}(x|\tf) \, , \quad \text{with} \quad V_{\mu}(x|\tf) \big|_{\tf=0} = U_\mu(x) \,,
    \label{eq:WF}
\end{equation}
where $\partial_{x, \mu}$ stands for the natural $\mathfrak{su}(3)$-valued differential operator with respect to the link variable $V_{\mu}(x|\tf)$ (see Appendix~A of Ref.~\cite{Luscher:2010iy}). 
One can show that expectation values of gauge-invariant observables at $\tf>0$ are renormalized quantities with a well-defined continuum limit~\cite{Luscher:2010iy,Luscher:2011bx}.

Since no particular differences w.r.t. the $\CPN$ model emerge in the MC analysis, to avoid repetitions, in the pure-gauge theory we focused mostly on the one dimensional MF analysis of flowed observables, where interesting considerations have to be made. By focusing on parity even observables, we consider the zero-momentum projection of the plaquette operator and the energy density at positive flow time $\tf$
\begin{equation}\label{eq:obs_tf}
\begin{aligned}
    \widetilde{\mathcal P}_{\tf}(x_0) &\equiv \frac{1}{L^3} \sum_{\vec{x}} \sum_{\mu > \nu} \frac{1}{18} \Re \Tr V_{\mu \nu}(x_0, \vec x|\tf) \, , \\
    \widetilde{\mathcal E}_{\tf} (x_0) &\equiv \frac{1}{L^3} \sum_{\vec{x}} \sum_{\mu > \nu} \frac14 G_{\mu \nu}^a(x_0, \vec x|\tf) G_{\mu \nu}^a(x_0,\vec x|\tf) \, , 
\end{aligned} 
\end{equation}
where $G_{\mu\nu}^a(x|\tf)$ and $V_{\mu\nu}(x|\tf)$ are the clover definition of the field-strength tensor and the plaquette, at positive flow time, respectively (see Refs.~\cite{Luscher:2010iy,Ce:2015qha}). The corresponding global observables are obtained by averaging over the coordinate $x_0$ and are denoted with $\mathcal{P}_{\tf}$ and $\mathcal{E}_{\tf}$ respectively.
Given their importance in scale setting and in monitoring the ergodicity of the Markov chain (see e.g.~\cite{Bruno:2014ova}), these observables are typically measured in all simulations, also with dynamical fermions, implying that our study may be readily tested and replicated in many state-of-the-art calculations.

\begin{figure}[ht]
	\centering
    \includegraphics[width=.45\textwidth,keepaspectratio]{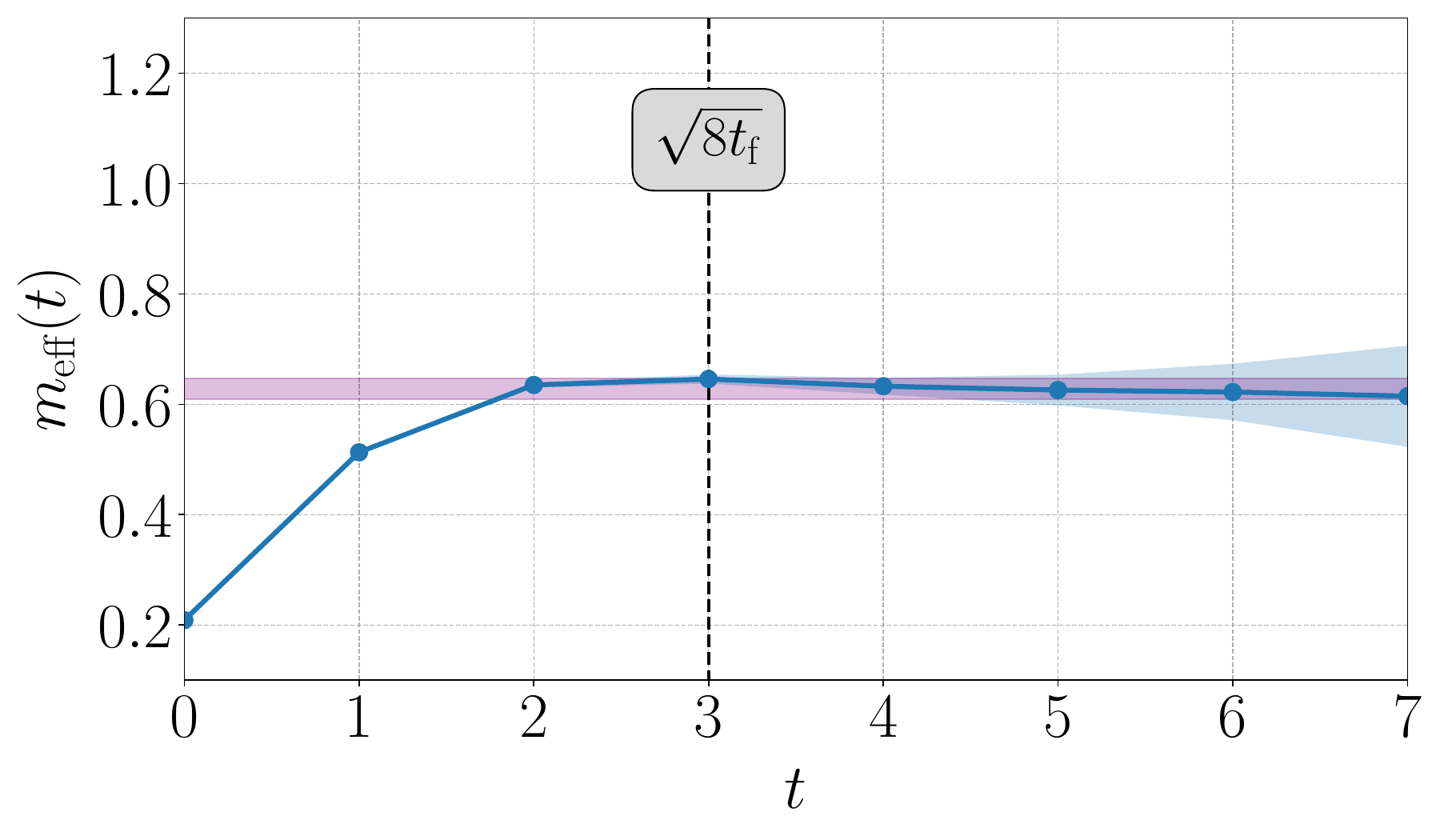}
    \includegraphics[width=.45\textwidth,keepaspectratio]{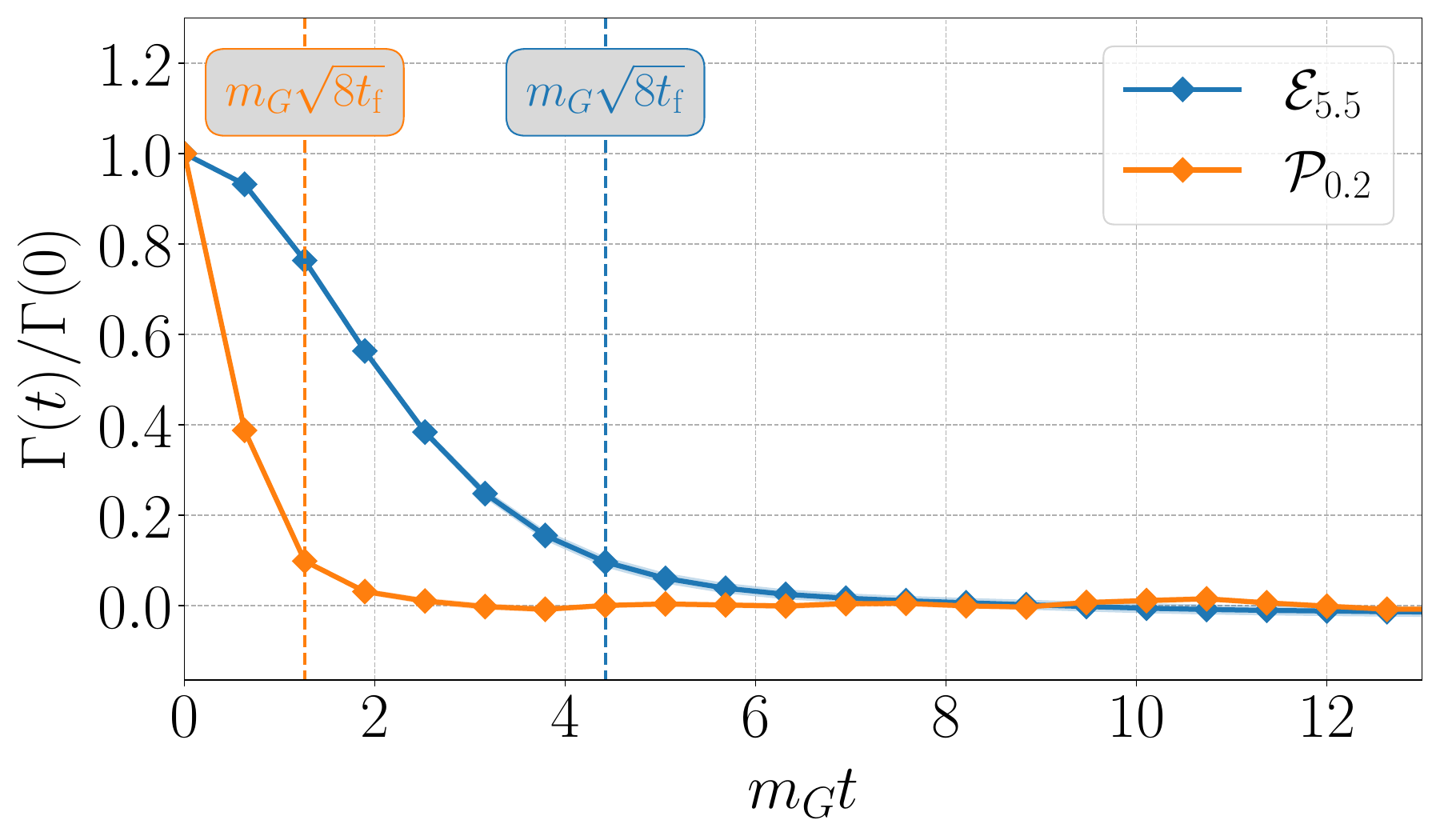}
    \caption{\textit{Left}: effective mass of the scalar glueball, in lattice units, from the plaquette two-point correlator smeared at $\tf = 1$. The purple horizontal band is taken from Ref.~\cite{Athenodorou:2020ani} and converted to lattice units as described in the main text. \textit{Right}: MF autocorrelation functions of the energy density and plaquette operators at two different flow times, $5.5$ and $0.2$ respectively. The former roughly corresponds to $t_0$.}
    \label{fig:smearing_SU3}
\end{figure}

\begin{figure}[ht]
	\centering
    \includegraphics[width=.45\textwidth,keepaspectratio]{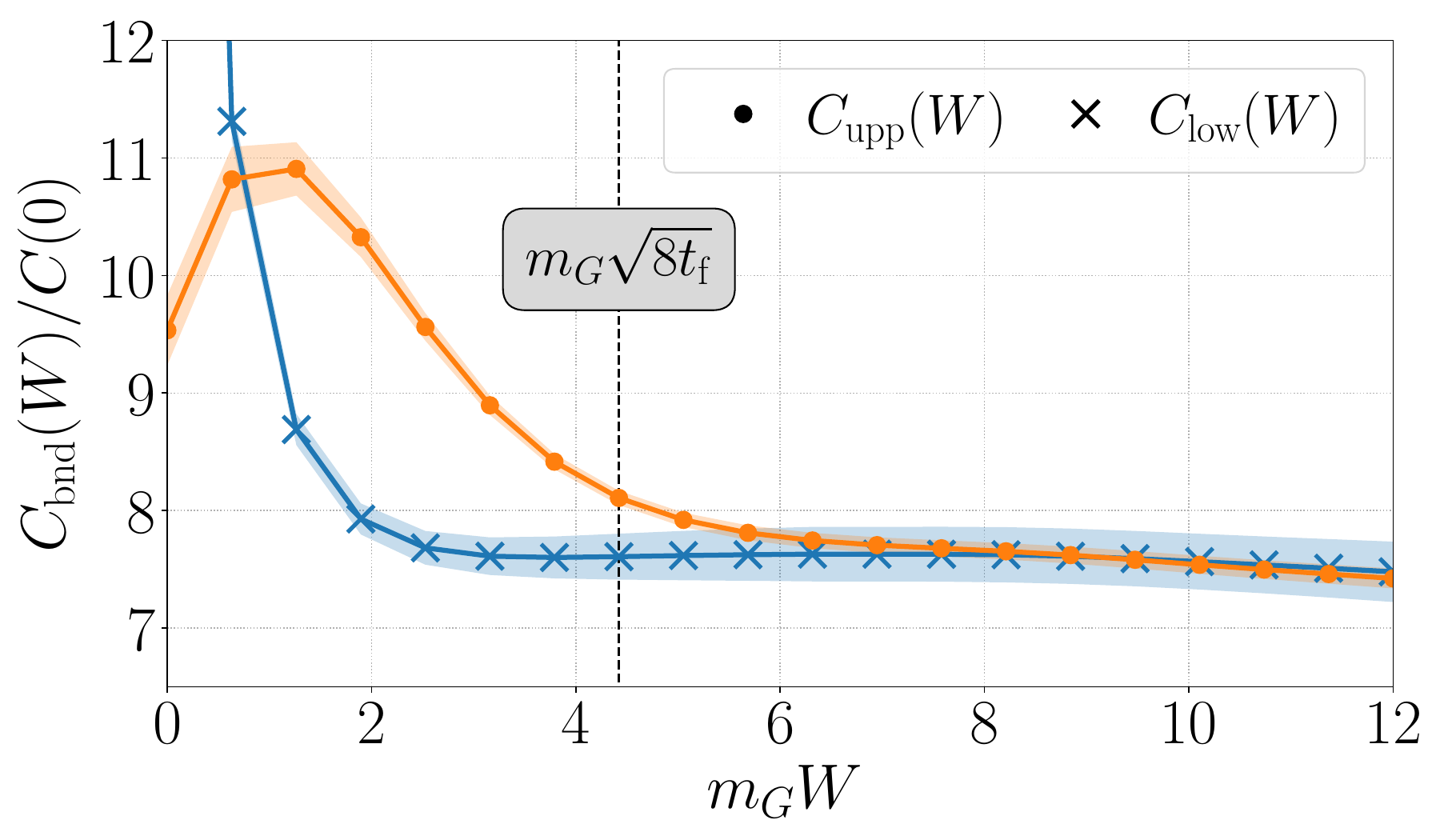}
    \includegraphics[width=.45\textwidth,keepaspectratio]{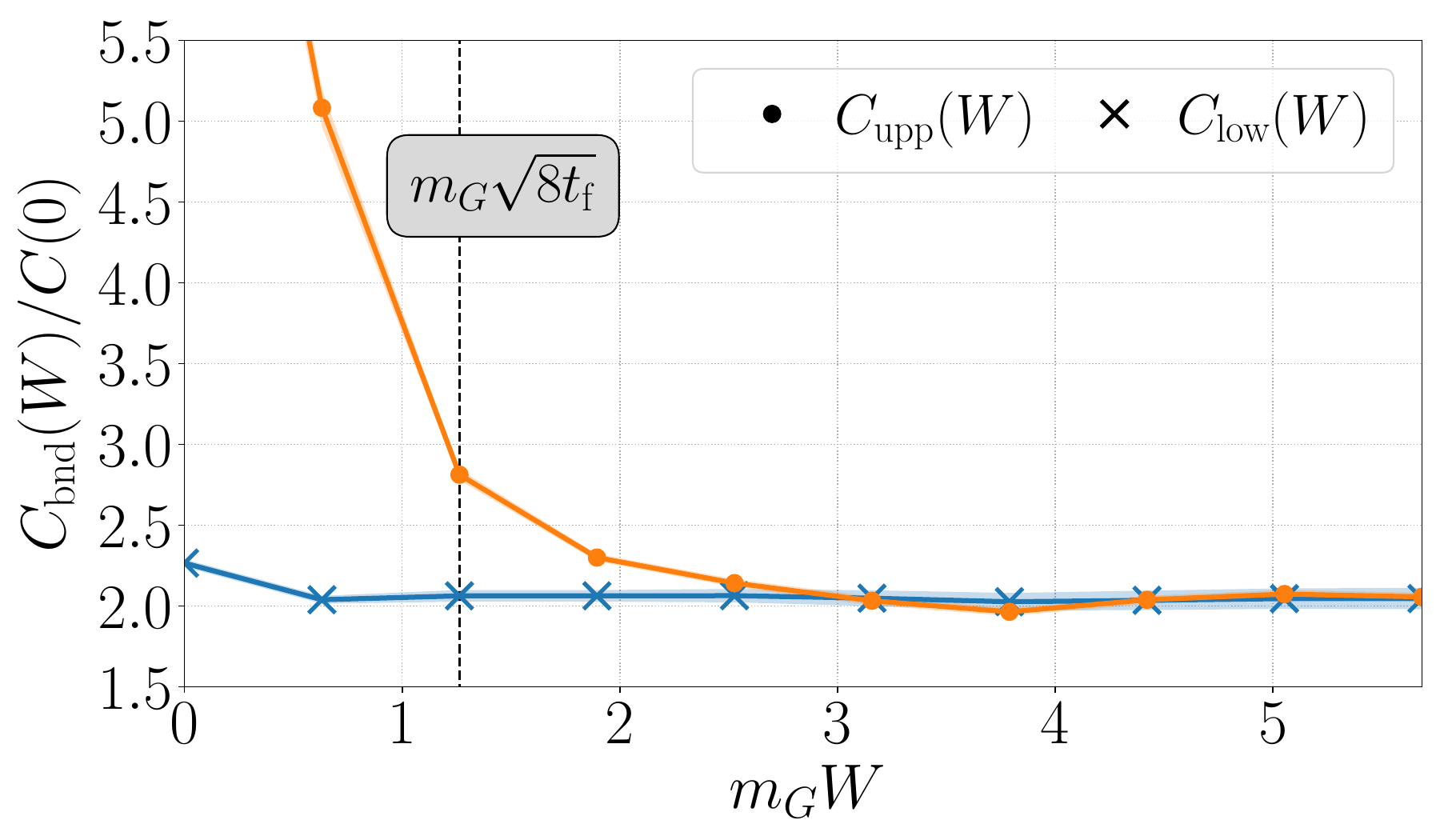}
    \caption{Bounding method applied to the MF autocorrelation functions of the observables $\mathcal{E}_{\tf}$ and $\mathcal{P}_{\tf}$ at different flowed times; the shaded bands represent the statistical error. In the upper bounds the value $\tau_0 = 4.75$ has been employed, which amounts to three times the (inverse) mass gap $\mG \sim 0.632$ extracted from the plateau of the effective mass in Fig.~\ref{fig:smearing_SU3}. \textit{Left}: $\mathcal{E}_{\tf}$ at $\tf=5.5$ in lattice units. \textit{Right}: $\mathcal{P}_{\tf}$ at $\tf=0.2$ in lattice units.}
    \label{fig:bounding_SU3}
\end{figure}

The Wilson flow is effectively a four-dimensional smearing procedure of the gauge links: composite operators built from flowed links are therefore non-local also along the direction used for the MF analysis, an effect spoiling the working hypothesis of the bounding method. Similarly to $G_4(t|x_0)$ in the $\CPN$ model, also here the representation in Eq.~\eqref{eq:gamma_exp} is not valid at short distances, but it is expected to be restored at separations larger than the smearing radius, approximately given by $\sqrt{8 \tf}$~\cite{Luscher:2010iy}. 
Therefore for $t \leq \sqrt{8 \tf}$ one should necessarily use the calculated autocorrelation function in the error computation and as a reminder of this fact, a vertical dashed line corresponding to $\sqrt{8 \tf}$ is always drawn in Figs.~\ref{fig:smearing_SU3} and \ref{fig:bounding_SU3}.

Our MF analysis has been performed on approx.~$82000$ MDUs collected from the simulations of a $32^3 \times 256$ lattice with $\beta = 6.17$ and using the HMC algorithm~\cite{Duane:1987de}. The product of the glueball mass with the time extent is approximately $160$ and the MF analysis has been carried out by skipping 600 MDUs (corresponding to approximately 6 times the autocorrelation time of the two observables in Eqs.~\eqref{eq:obs_tf} at flow time equal to $t_0$), leading to $O(130)$ independent fields.
We have measured the standard Wilson flow scale $t_0$ defined in Ref.~\cite{Luscher:2010iy} obtaining the value $\sqrt{t_0} = 2.34447(72)$ (in lattice units), compatible with other estimates from the literature~\cite{Luscher:2010iy}. Making use of the physical value $\sqrt{t_{0,\mathrm{phys}}} = 0.176(4) \text{~fm}$ from Ref.~\cite{Ce:2015qha} we determine the approximate inverse lattice spacing in physical units, $a^{-1} \simeq 2.63 \text{~GeV}$.
We extract the mass gap of the theory, in our case the $0^{++}$ glueball mass $\mG$, from the asymptotic behavior of the two-point function of the plaquette operator at positive flow time. The result for the effective mass is reported in the left panel of Fig.~\ref{fig:smearing_SU3} and is in good agreement with a recent determination taken from Ref.~\cite{Athenodorou:2020ani}, converted to lattice units using the numbers above.
Since the autocorrelation function of the plaquette coincides with its connected two-point correlator we can study the MF analysis rather accurately, thanks to our large statistics. At short distances, the effect of smearing is visible   from the right panel of Fig.~\ref{fig:smearing_SU3} where we compare the normalized autocorrelation function of the plaquette and energy density at very different flow times (as in the previous study, the error of autocorrelation functions has been estimated along the Monte Carlo direction). 

To demonstrate in practice the limits of applicability of the bounding method we calculate lower and upper bounds at all separations for the two different smeared scenarios and report the corresponding results in the two panels of Fig.~\ref{fig:bounding_SU3}. The two chosen examples are for $\mathcal{E}_{\tf}$ with $\tf \simeq t_0$ (left panel) and $\mathcal{P}_{\tf}$ (right panel) with $\tf=0.2$ in lattice units. For the latter no particular differences w.r.t. the previous analysis emerged, even at $t \leq \sqrt{8 \tf}$, while for the former the failure of the bounding method manifests itself via an inverted hierarchy at shorter distances, which is later restored for values of $W$ comparable with the smearing radius. In practice we find that the bounding method becomes applicable at distances of $O(\sqrt{8 \tf})$ where the locality of the flowed operators is effectively restored. 

\section{Conclusions}

Statistical error estimation in lattice field theory simulations is a fundamental and central step for every calculation. The presence of autocorrelations requires the knowledge and summation of the autocorrelation function, which in practice is truncated to a fixed window. Automatic windowing procedures are very important, especially in modern calculations, which involve complicated manipulations of primary observables (the correlation functions of the theory) to extract physically relevant quantities. Various practical truncation schemes have been studied in the literature \cite{Madras:1988ei, Wolff:2003sm, Schaefer:2010hu} and in this work we have proposed an alternative method, based on the definition of strict lower and upper bounds of the autocorrelation function. The latter was already examined in Ref.~\cite{Schaefer:2010hu} and in our work we re-assess it by complementing it with a lower bound, which naturally leads to the formulation of a windowing procedure based on their saturation.
While our method finds its most natural application in the context of error analysis based on the master-field approach, we have also discussed its application in the context of Monte Carlo data; in this case the properties of the algorithm matter, specifically detailed balance and ergodicity.
Our manuscript is explicitly targeted towards the so-called $\Gamma$-method, but it could be used in conjunction with resampling strategies based on binned data, as a tool to identify a proper bin size.
For this study we have been inspired by the recent progress in the context of the muon anomalous magnetic moment, where the vector two-point correlator in QCD is summed with specific weight functions to reproduce the Hadronic-Vacuum-Polarization contribution to $(g-2)_\mu$. Similarly to the autocorrelation function it suffers from a signal-to-noise problem and requires a truncation, which we investigated here in the context of error analysis.

In this work we tested the idea on both synthetic and real data taken from Monte Carlo simulations, based on the HMC algorithm, of two simple toy models, the $\CPN$ theory in two dimensions and the $\mathrm{SU}(3)$ pure-gauge theory in four. 
In the MC analysics, we found that the bounding method is stable and consequently we expect that it could be used in more costly simulations of QCD as well, provided that the algorithms employed satisfy the necessary properties listed above. For those cases when the slowest autocorrelation mode is not known a priori, we have studied a few possibilities: an iterative procedure which starts from a guess extracted from the lower bound, a more refined version based on a calculable estimate of $\tau_0$ and the more traditional spectral analysis of the autocorrelation function based on the GEVP. For the former an intrinsic dependence on tunable parameters is still present, making it very similar in spirit to Wolff's procedure~\cite{Wolff:2003sm} and its variations~\cite{Schaefer:2010hu}. For the latter, evidently, one needs a few observables sufficiently coupled to the low modes of the Markov chain and capable to cover the relevant subset of $\tau_n$'s.
In our numerical tests we found that, when this is the case, the GEVP approach provides a good guidance in the definition of the upper bound and its application to the MC analysis of state-of-the-art Lattice QCD simulations remains an interesting open question to be further pursued. Nevertheless we stress that for pathological simulations suffering from critical slowing down, improved algorithms and long Markov chains are in any case needed.

In our manuscript we have also examined the bounding method in the case of one-dimensional MF analysis, where its working hypothesis are  satisfied. In this case the autocorrelation function is a correlator of the theory and its long-distance behavior is governed by (at least) the mass gap of the theory. In practice and depending on the nature of the observable, heavier states may dominate and in our investigation of the $\CPN$ model we verified this fact, showing that the upper bound based on the mass gap is a fairly safe choice. Provided that a sufficiently long lattice is available, the bounding method can be implemented straightforwardly for local observables, and from a certain minimal distance also for non-local observables, such as two-point correlators with non-zero physical separation or smeared operators with a significant footprint. We tested the bounding method for these two scenarios in the $\CPN$ model and in the pure-gauge theory respectively. For the latter, we examined observables smeared in all four-directions using the gradient flow, a standard for the community. We checked that the bounding method can be applied at distances larger than either the physical separation of the two operators or the smearing radius, in the two cases respectively. The case of WF observables is particularly relevant in the presence of dynamical light fermions: for example smearing up to $t_0$ introduces a footprint of approximately $\sqrt {8 t_0} \sim 0.5 ~\mathrm{fm}$ which is only half of the physical pion wavelength and therefore a stronger signal is expected compared to our tests in the pure-gauge theory, where the autocorrelation function falls off at a faster rate.

Finally we observed that in general, the lower bound develops long plateaus which provide fairly good estimates of the error by themselves; an alternative strategy to the saturation of the bounds proposed here, could in fact be based on the sole stability of the lower bound at large windows, which would be completely data-driven.

\acknowledgments

We thank the lattice group of the University of Milano-Bicocca for stimulating discussions and M. N. Colombo for crosschecks on the code for the $\mathbb{CP}^{M-1}$ model. Special thanks go to M.~Dalla Brida for helpful and insightful discussions on Monte Carlo methods. G.M. is grateful to F.A.~Bresciani for several valuable advices and interesting discussions on the topic. The authors would like to thank the RBC/UKQCD collaboration for useful discussions on related topics and ongoing collaborations. 
Additionally M.B. would like to thank M.~C\'e, A.~Francis, P.~Fritzsch, J.~Green, M. Hansen and A.~Rago for fruitful conversations on topics related to master-field simulations. At the beginning of the project, M.B. was supported by the national program for young researchers ``Rita Levi Montalcini''. 
This work is (partially) supported by ICSC - Centro Nazionale di Ricerca in High Performance Computing, Big Data and Quantum Computing, funded by European Union – NextGenerationEU.

\appendix

\section{Spectral decomposition of the autocorrelation function}
\label{app:gamma}

In this Appendix we sketch the spectral decomposition of the autocorrelation function for the MC case by repeating some of the arguments presented in Refs.~\cite{Schaefer:2010hu, Luscher:2010ae, Seabrook_2023} (and references therein).

We denote with $q$ the configurations (or states) generated by the Markov process and with $\mathcal M(q' \leftarrow q)$ the probability of transitioning from configuration $q$ to configuration $q'$. The equilibrium distribution $P(q)$ is an eigenvector of $\mathcal M$ with unit eigenvalue, namely $\mathcal M \otimes P= P$, where the symbol $\otimes$ denotes the sum over all possible configurations or states. 
By introducing the redundant notation $\obs_\alpha(q)$, with the explicit dependence on the configuration $q$ (relevant only in this Appendix), its $t$-dependence is replaced according to
\begin{equation}
    \obs_\alpha(t) \to \mathcal M^t \otimes \obs_\alpha \,, \quad \text{with} \quad \mathcal M^{\,t} \equiv \overbrace{\mathcal M \otimes \cdots \otimes \mathcal M}^{t}
\end{equation}
and expectation values can be re-written as $\langle \obs_\alpha \rangle = \obs_\alpha \otimes P$.
Thanks to the operator
\begin{equation}\label{eq:T_qqp}
    \mathcal T(q' \leftarrow q) = P(q')^{-1/2} \mathcal M(q' \leftarrow q) P(q)^{1/2} 
\end{equation}
and the weighted fluctuations $\Delta \obs_\alpha(q) = \delta \obs_\alpha(q) P(q)^{1/2}$,
the autocorrelation function in Eq.~\eqref{eq:gamma_def} for a Markov chain is rewritten as
\begin{equation}
    \Gamma_{\alpha\beta}(t) = \Delta \obs_\alpha \otimes \mathcal T^{\,t} \otimes \Delta \obs_\beta \,, \quad \mathcal T^{\,t} \equiv \overbrace{\mathcal T \otimes \cdots \otimes \mathcal T}^{t} \,.
    \label{eq:gamma_T}
\end{equation}
When the underlying algorithm respects detailed balance, i.e.
\begin{equation}
    \mathcal M(q' \leftarrow q) P(q) = \mathcal M(q \leftarrow q') P(q') \,,
\end{equation}
and is ergodic, the operator $\mathcal T$ is symmetric and has real eigenvalues $|\lambda_n|<1$ with eigenvectors $u_n(q)$, in addition to the unit eigenvalue inhereted from $\mathcal M$ and defined by $\mathcal T \otimes P^{1/2} = P^{1/2}$.
Without loss of generality we take $\lambda_0$ as the largest eigenvalue and the ordering $|\lambda_{n-1}| \geq |\lambda_n|$.
With little algebra one can show that 
\begin{equation}
    \Gamma_{\alpha\beta}(t) = \sum_n c_{\alpha\beta}^n (\lambda_n)^t \,, \quad \text{with} \quad
    c_{\alpha\beta}^n = \big[ \Delta \obs_\alpha \otimes u_n\big] \big[u_n \otimes \Delta \obs_\beta \big] \,,
    \label{eq:gamma_spec_dec}
\end{equation}
such that the positivity of $c_{\alpha\alpha}^n$ easily follows.
Notice that being a connected correlator, the stationary mode corresponding the eigenvalue equal to 1 is excluded from the spectral decomposition of the autocorrelation function. At asymptotically large times the autocorrelation function decays exponentially, namely $\lambda_0=e^{-1/\tau_0}>0$ and $\tau_0$ is typically called the exponential autocorrelation time $\tau_\mathrm{exp}$. At this stage, despite the positivity of $c_{\alpha\alpha}^n$, we are not yet in the position to derive strict bounds, since the eigenvalues may be negative, $-1 < \lambda_n < 1$. However, when the autocorrelation function is expressed in terms of the positive-definite operator $\mathcal T^{\,T} \otimes \mathcal T$, the representation in Eq.~\eqref{eq:gamma_exp} is automatically valid. In practice this may be achieved by measuring every other (or in general by skipping an even number of) Markov step.

In the MF analysis, well-defined theories with a mass gap have a bounded transition matrix defined in terms of a positive-definite Hamiltonian: by working in its eigen basis the contribution of $(\lambda_n)^t$ is mapped to the typical (Euclidean) exponential factors $e^{-E_n t}$, with $E_n$ the spectrum of the system, and the coefficients $c_{\alpha\beta}^n$ are matrix elements between the vacuum and the $n$-th Hamiltonian eigenstate $|n\rangle$.

\section{MC and MF simultaneous analysis} 
\label{app:MC_MF}

\begin{figure}[ht]
	\centering
    \includegraphics[width=.5\textwidth,keepaspectratio]{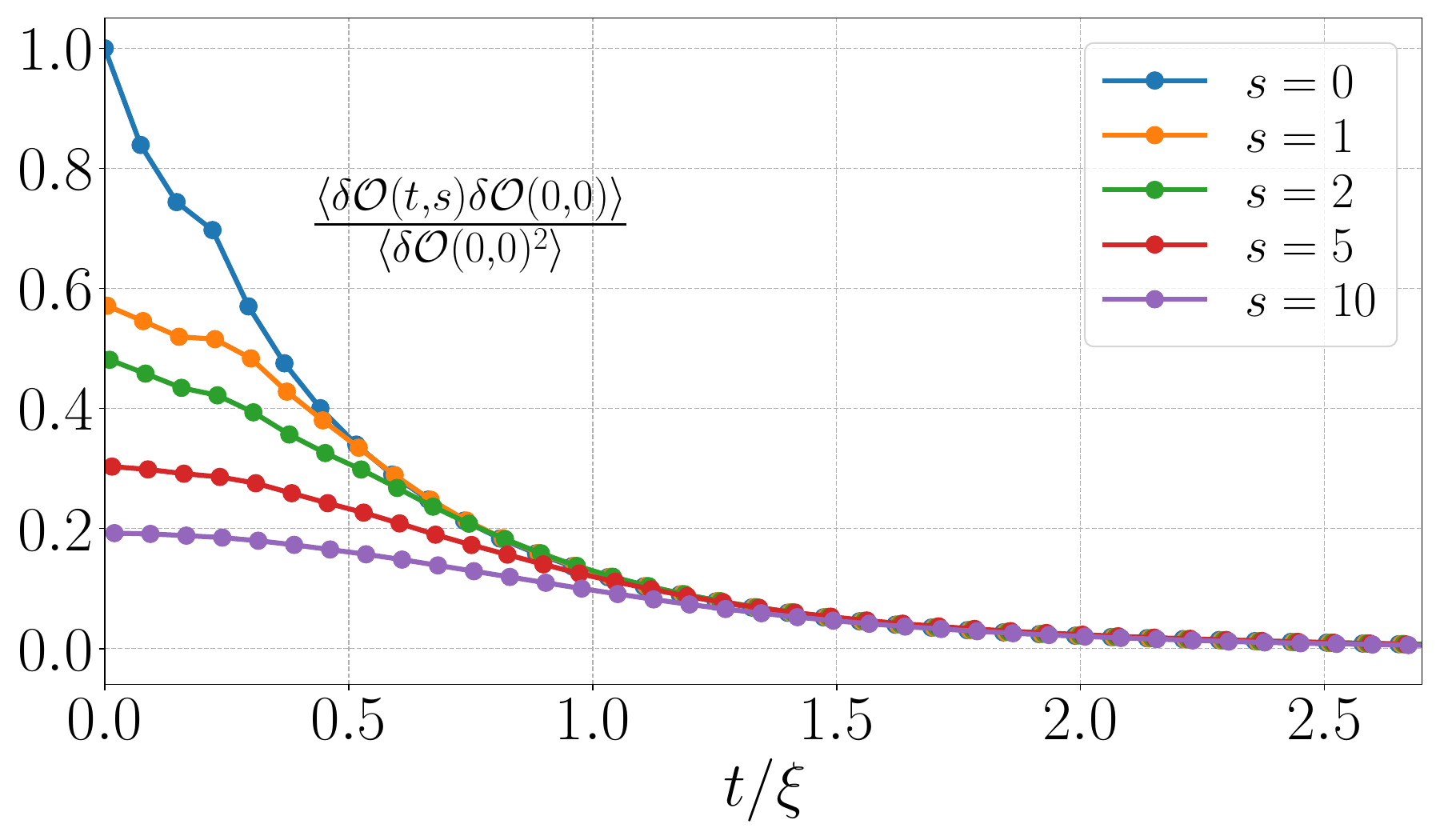}
    \caption{Autocorrelation function along the MF and MC directions, labeled by $t$ and $s$ respectively. The chosen observable corresponds to $G_2(x_0=3)$, as one can infer from the discontinuity at $t=x_0=3$. At large $s$, but fixed $t$, the autocorrelation function decays exponentially as expected, but in addition we observe that it becomes smoother.}
    \label{fig:gamma_MC_MF}
\end{figure}

When several configurations of sufficiently large lattices are available, one may be tempted to consider the superior autocorrelation function
\begin{equation}
    \langle \delta \obs_\alpha(t, s) \, \delta \obs_\beta(0,0) \rangle \,,
    \label{eq:Gamma_ts}
\end{equation}
where the two arguments $t,s$ in $\obs_\alpha(t,s)$ denote the two separate analysis directions, MF and MC. By truncating the two-dimensional integral, a reduced error of the error and consequently a better stability is in order. This has been explored in Ref~\cite{RBC:2023pvn} where saturation along both directions has been achieved from numerical inspection and safe summation windows. Introducing a rigorous bounding procedure requires a ``first-principle'' knowledge of the function above, and for that one could start from the results of Ref.~\cite{Luscher:2011qa}, which however show the non-renormalizability of the HMC algorithm. Using our high-statistics sample of the $\CPN$ model at $\beta=0.95$ we calculated the autocorrelation function in Eq.~\eqref{eq:Gamma_ts} for $\obs_\alpha=\obs_\beta = G_2(x_0=3)$ along the MF and MC directions. Contrary to the results presented in the main text where we considered practically independent fields, here we measured the observable on every trajectory, each 1 MDU long. MC autocorrelations are quite sizeable as one can observe from the results reported in Fig.~\ref{fig:gamma_MC_MF}.

Defining a suitable summation procedure for this case goes beyond the scope of this paper and is deferred to future studies. Here we simply note that by interpreting MC time as an additional direction in a five dimensional theory~\cite{Luscher:2011qa}, the autocorrelation function along the MF direction but at non-zero MC time may be explained in terms of smeared operators, in analogy to the WF scenario. This observation is supported by the curves shown in Fig.~\ref{fig:gamma_MC_MF}, where the discontinuity at $t=x_0$ is smeared out for $s \neq 0$. In practical applications where the original fluctuations $\delta \obs_\alpha$ are binned, this effect may force the applicability of the bounding method to longer distances, similarly to the case of the WF.

\section{The spectrum of the $\CPN$ model}
\label{app:cpn}

The $\CPN$ model admits a local $\mathrm{U}(1)$ symmetry, manifest from the presence of the $\lambda_\mu(x)$ field in the action, and a global $\mathrm{SU}(M)$ symmetry.
During the 80's it attracted significant attention thanks to its similarities with QCD and the analytic results one can obtain through $1/M$ expansions (see e.g.~\cite{Witten:1978bc,DAdda:1978vbw}). Using these techniques the lowest bound states of the spectrum have been studied in Refs.~\cite{Witten:1978bc,Haber:1980uy}, finding that the mass gap of the theory is given by a positive parity state in the adjoint representation of $\mathrm{SU}(M)$, followed by a singlet. Negative parity states, both in the adjoint and singlet representations, appear instead at higher scales and are expected to be degenerate. For an earlier numerical investigation see also Ref.~\cite{IRVING1992521}.

The generating functional for connected correlators involving the group and gauge invariant operators in the adjoint representation $P^a(x) = \Tr[P(x) t^a]$ is defined by
\begin{equation}
    \mathcal Z[J] =  \int Dz^\dagger Dz \, D\lambda \, e^{-S[z^\dagger \! ,z,\lambda]} \, \exp\left\{- \int \dd[2]{x} \, J_a(x) P^a(x) \right\} \,,
\end{equation}
where the matrices $t^a$ are the $M^2-1$ generators of $\mathrm{SU}(M)$, normalized such that $\Tr(t^a t^b) = \frac12 \delta^{ab}$. The parity-even two-point correlator where the lowest bound state is expected to propagate at long distances is therefore obtained from \cite{IRVING1992521, Rindlisbacher:2016cpj}
\begin{equation}
    G_A(x,y) = \sum_{a=1}^{M^2-1} \left. \frac{\partial^2 \log \mathcal Z[J]}{\partial J_a(x) \partial J_a(y)} \right|_{J=0}= \sum_{a=1}^{M^2-1} \Big\{ \langle P^a(x) P^a(y) \rangle - \langle P^a(x) \rangle \langle P^a(y) \rangle \Big\} \,,
\end{equation}
where we have summed all ``flavors''. Using the following Fierz identity for $\mathrm{SU}(M)$ generators in the defining representation
\begin{equation}
    \sum_{a=1}^{M^2-1} [t^a]_{ij} [t^a]_{kl} = \frac12 \left( \delta_{il} \delta_{jk} - \frac{1}{M} \delta_{ij} \delta_{kl} \right)
\end{equation}
together with $z^\dagger(x) z(x) = \Tr P(x) = 1$ one finds, with little algebra,
\begin{equation}
    2 G_A(x,y) = \Tr \langle P(x) P(y) \rangle - \frac{1}{M} \,,
\end{equation}
which coincides with $G_2(x_0)$ once projected to zero spatial momentum. Provided that $\langle P^a(x) \rangle$ vanishes, one finds the additional relation
\begin{equation}
    \sum_a [t^a]_{ij} \langle P^a(x) \rangle = 0 \quad \to \quad \langle P_{ij}(x) \rangle = \frac{\delta_{ij}}{M} \,.
\end{equation}

Now we consider a two-point connected correlation function built from composite operators of the form
\begin{equation}
    \Tr \big[P(x_1) P(x_2) \dots P(x_n) \big] \,,
\end{equation}
where the points $x_1\dots x_n$ are clustered together within a small region and the two operators are displaced in two non-overlapping areas. Individually they transform as singlets under $\mathrm{SU}(M)$, and as a consequence the propagating states belong to the singlet representation, which according to Refs.~\cite{Haber:1980uy} are expected to be more energetic. This is the case for the correlation function $G_4(t|x_0)$ that governs the error of $G_2(x_0)$, a fact that is reflected by our numerical results. A precise numerical check goes beyond the scope of this paper and here we simply note that the extracted mass is approximately twice as large as the adjoint. 

\section{Additional examples in the $\CPN$ model}
\label{app:obs_cpn}

\begin{figure}[ht]
	\centering	
    \includegraphics[width=.45\textwidth,keepaspectratio]{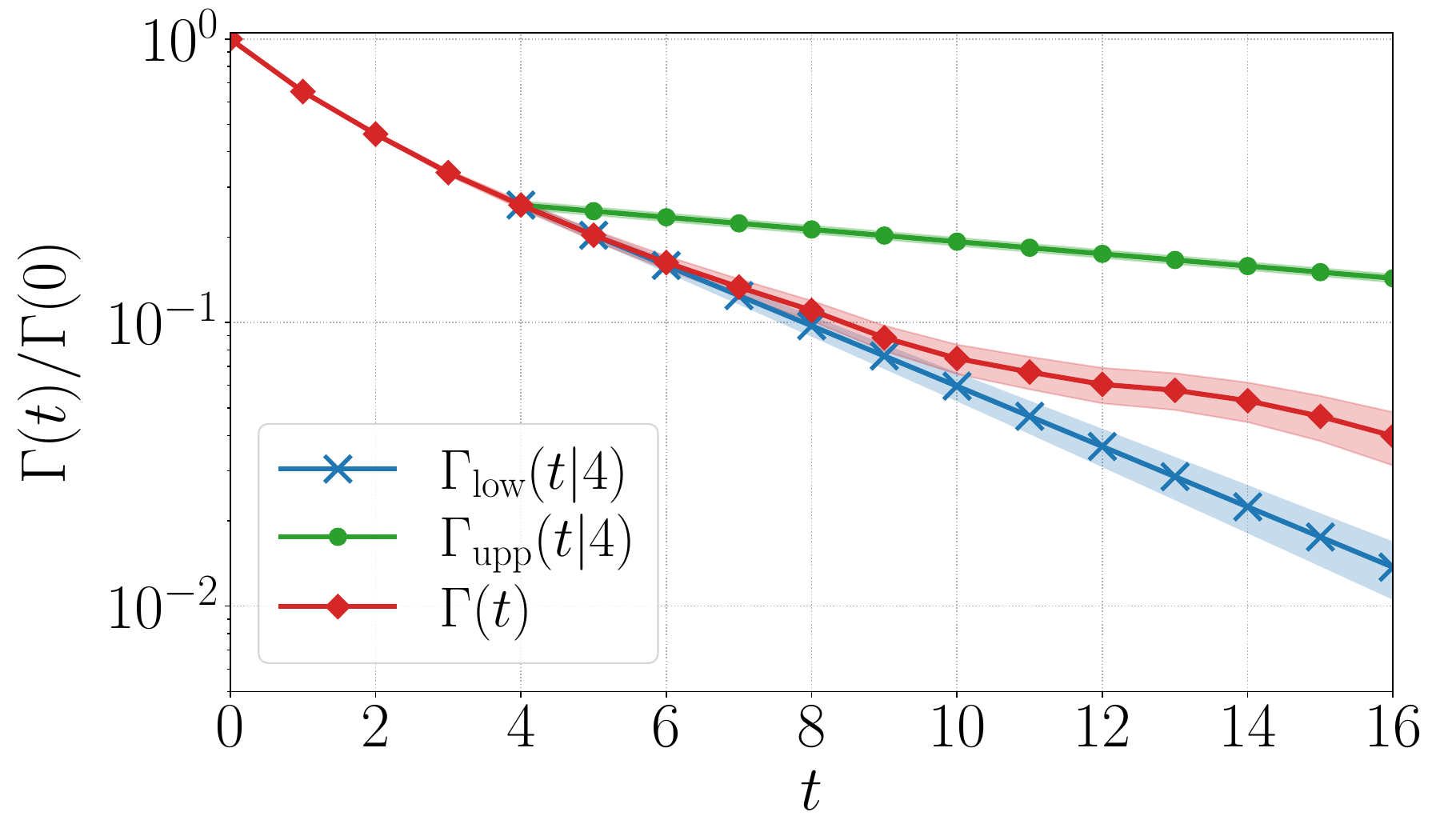}
    \includegraphics[width=.45\textwidth,keepaspectratio]{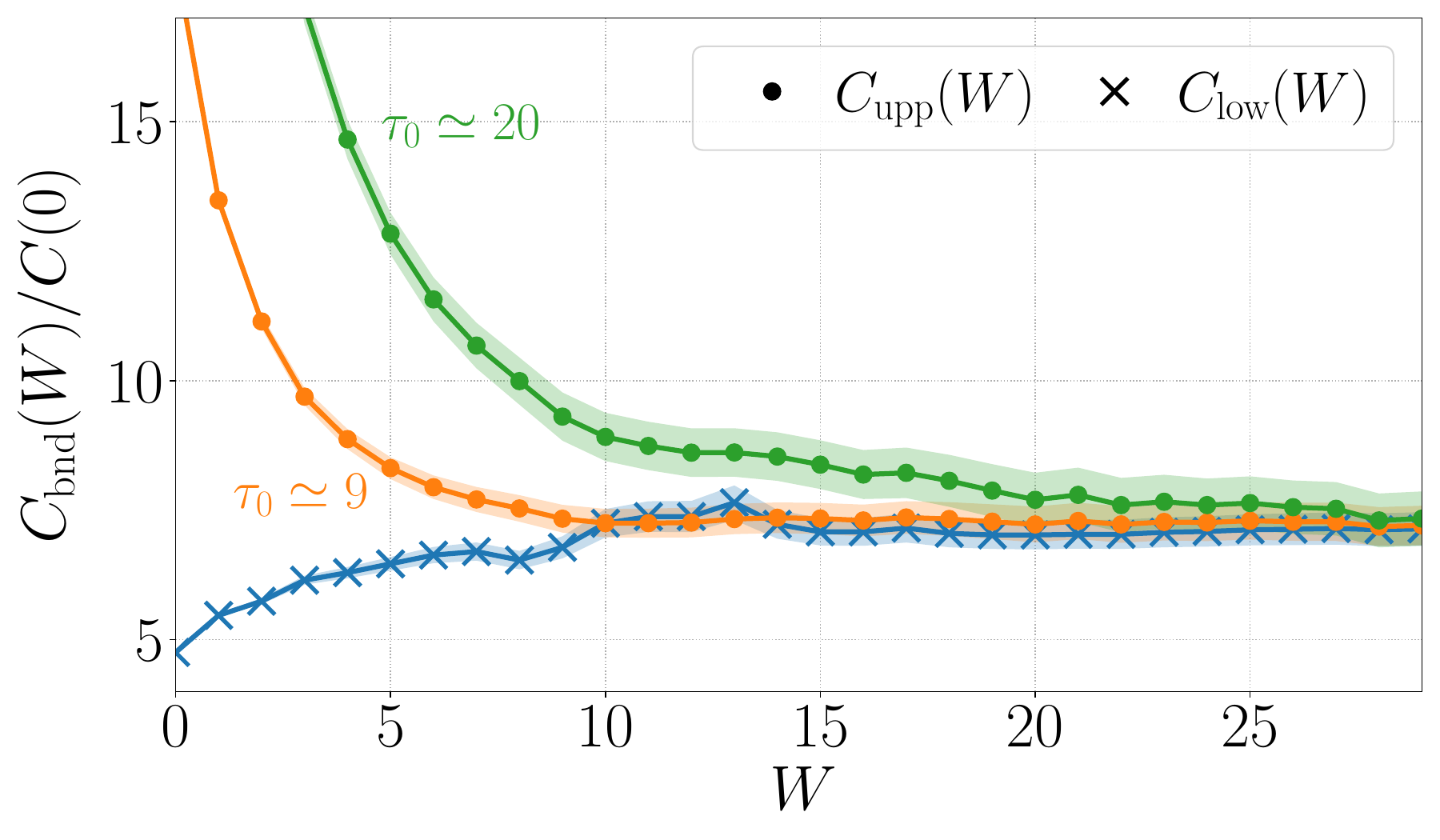} \\
    \includegraphics[width=.45\textwidth,keepaspectratio]{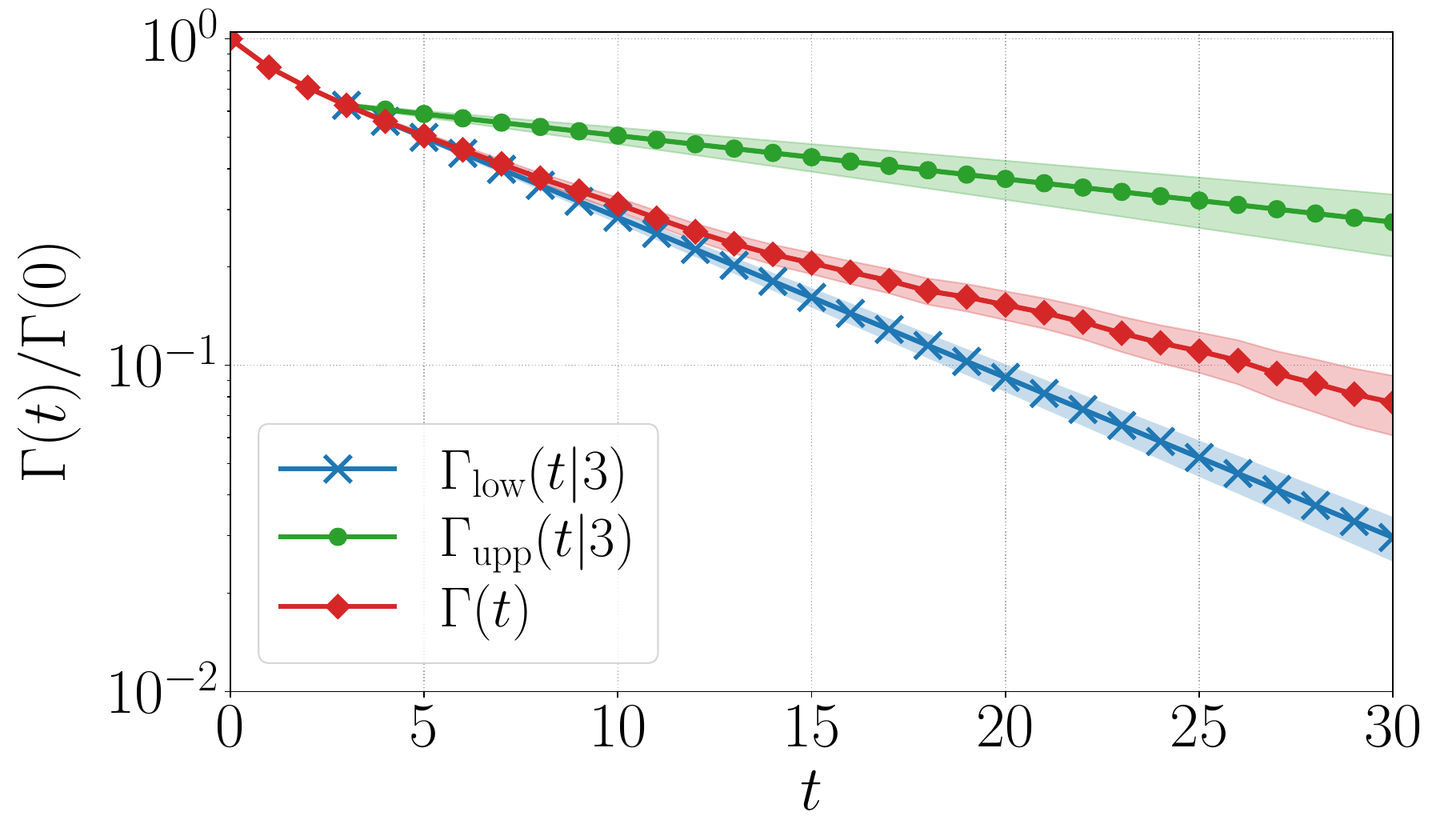}
    \includegraphics[width=.45\textwidth,keepaspectratio]{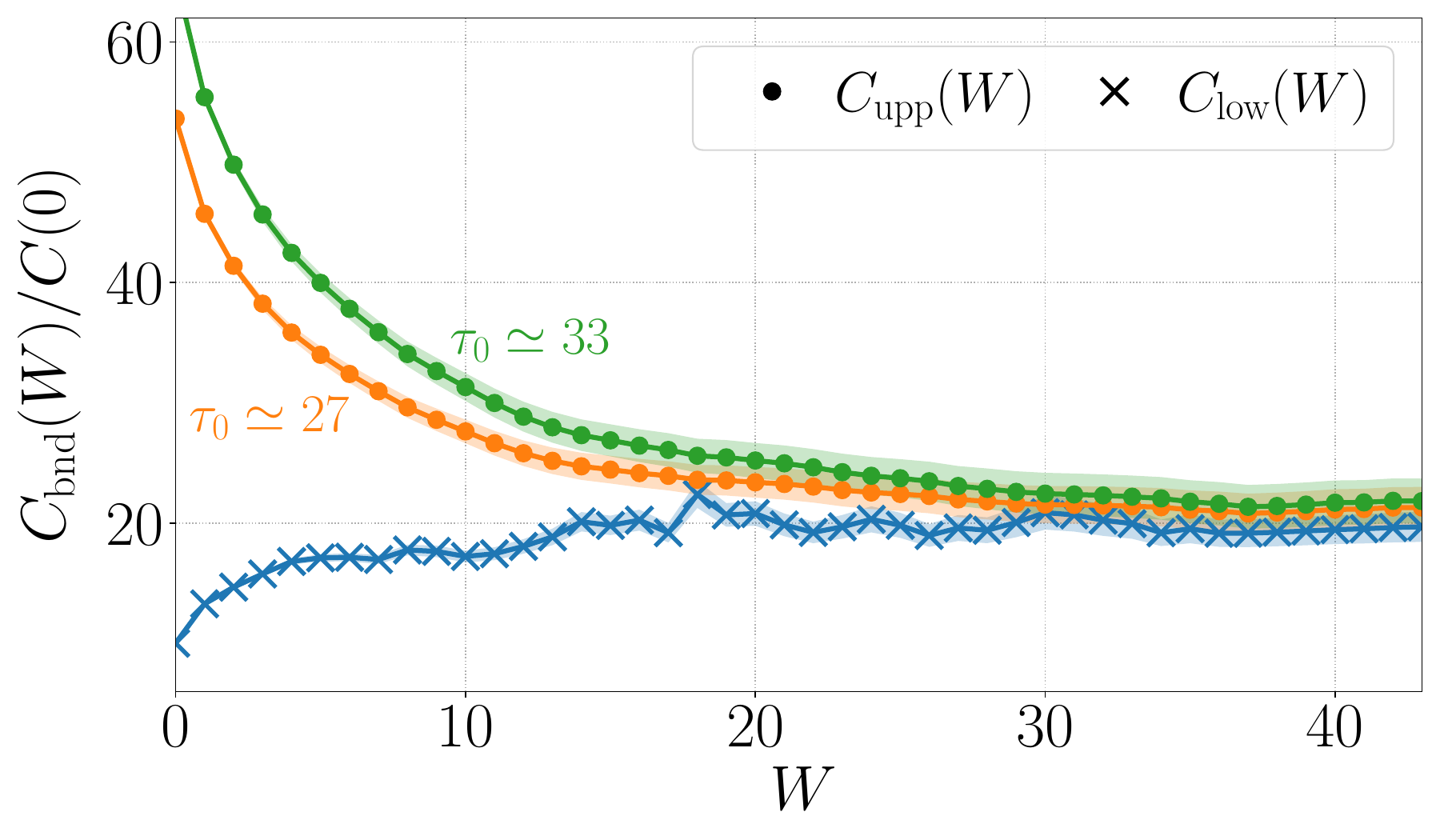}
    \caption{Example of application of the Bounding method to the MC autocorrelation function of $E$ (top) and $\chi_\mathrm{M}$ (bottom). \textit{Left}: autocorrelation function and its bounds from Eq.\eqref{eq:bounds} using $W = 3$. The last iteration of $\tau_0$ in the corresponding right panel has used for the calculation of $\Gamma_\mathrm{upp}$; the shaded bands represent the statistical errors. \textit{Right}: integrals $C_\mathrm{upp}(W)$ and $C_\mathrm{low}(W)$ as a function of $W$. Two representative upper bounds are shown in orange and green respectively (see subsection~\ref{subsec:cpn}).}
    \label{fig:Bnd_MC_CPN_2}
\end{figure}

In this appendix we collect plots for the observables $E$ and $\chi_\mathrm{M}$, analogous to the ones in Fig.~\ref{fig:Bnd_MC_CPN}. The same procedure discussed in subsection~\ref{subsec:cpn} has been employed to calculate their error and in the right panels of the Fig.~\ref{fig:Bnd_MC_CPN_2} we plot a few representative examples of the curves $C_\mathrm{upp}(W)$.

\section{Details on the GEVP implementation}
\label{app:GEVP_mel_en}

In this appendix we collect relevant formulae used in the calculation of energies and matrix elements from the solution of the GEVP in Eq. \eqref{eq:GEVP_gamma}.
The GEVP can be solved at every time-slice $t$ at fixed $t_0$ or by spanning both $(t, t_0)$ while taking the difference $t-t_0$ constant and properties of individual states are extracted for $t,t_0 \gg 0$. Energies are obtained from the following approximation
\begin{equation}
    E_n(t,t_0) \simeq -\frac{1}{t-t_0} \log \lambda_n(t,t_0) \, , 
    \label{eq:ground_single}
\end{equation}
assuming $t>t_0$ and $t,t_0 \gg 1$ (depending on the error analysis, the time coordinate refers to either Monte-Carlo time or physical time), while the coefficients $c^n_{\alpha\beta}$ in Eq.~\eqref{eq:gamma_exp} can be extracted from the effective matrix elements (we suppress the dependence on $t, t_0$ of energies and eigenvectors)
\begin{equation}\label{eq:GEVP_coeffs}
    c_{\alpha \beta}^n \simeq \frac{e^{E_n t} \big[ \sum_\gamma v_{\gamma n} \Gamma_{\gamma \alpha}(t) \big] \big[ \sum_\rho \Gamma_{\beta \rho}(t) v_{\rho n} \big]}{ \sum_{\gamma \rho} v_{\gamma n} \Gamma_{\gamma \rho}(t) v_{\rho n}} \, .
\end{equation}

The numerical stability of the solution is often compromised by a low number of measurements and by statistical fluctuations in the data. For this reason, we adopt a projection onto the subspace of the most relevant eigenvectors of a given $\Gamma(t)$. The interested reader may consult Refs.~\cite{Bulava:2024jpj,RBC:2023xqv} for alternative procedures and Ref.~\cite{RBC:2024fic} for an actual application.
What we call \emph{projected} GEVP (pGEVP) in the main text is defined from the following steps:
\begin{enumerate}
    \item at fixed $t_P$ we diagonalize the matrix $\Gamma_{\alpha\beta}(t_P)$ and denote with $l_n(t_P)$ and $u_n(t_P)$ ($n = 1, \ldots, N_\mathrm{obs}$) the eigenvalues and eigenvectors respectively;
    \item we select the most relevant $N_P$ eigenvectors by checking that the corresponding eigenvalues are statistically significant and different from zero;
    \item using the selected eigenvectors labeled by the indices $k_1, k_2, \ldots, k_{N_P}$ we construct the $N_\mathrm{obs} \times N_P$ projection matrix
    \begin{equation}
        \mathbb{P}(t_P) = [u_{k_1}(t_P), u_{k_2}(t_P), \ldots, u_{k_{N_P}}(t_P)] \,;
    \end{equation}
    \item we calculate the projected $N_P \times N_P$ correlator matrix $\widetilde{\Gamma}(t;t_\mathrm{P}) \equiv \mathbb{P}^T(t_\mathrm{P}) \Gamma(t) \, \mathbb{P}(t_\mathrm{P})$ and solve the corresponding GEVP.
\end{enumerate}

\begin{figure}[ht]
	\centering
    \includegraphics[width=.45\textwidth,keepaspectratio]{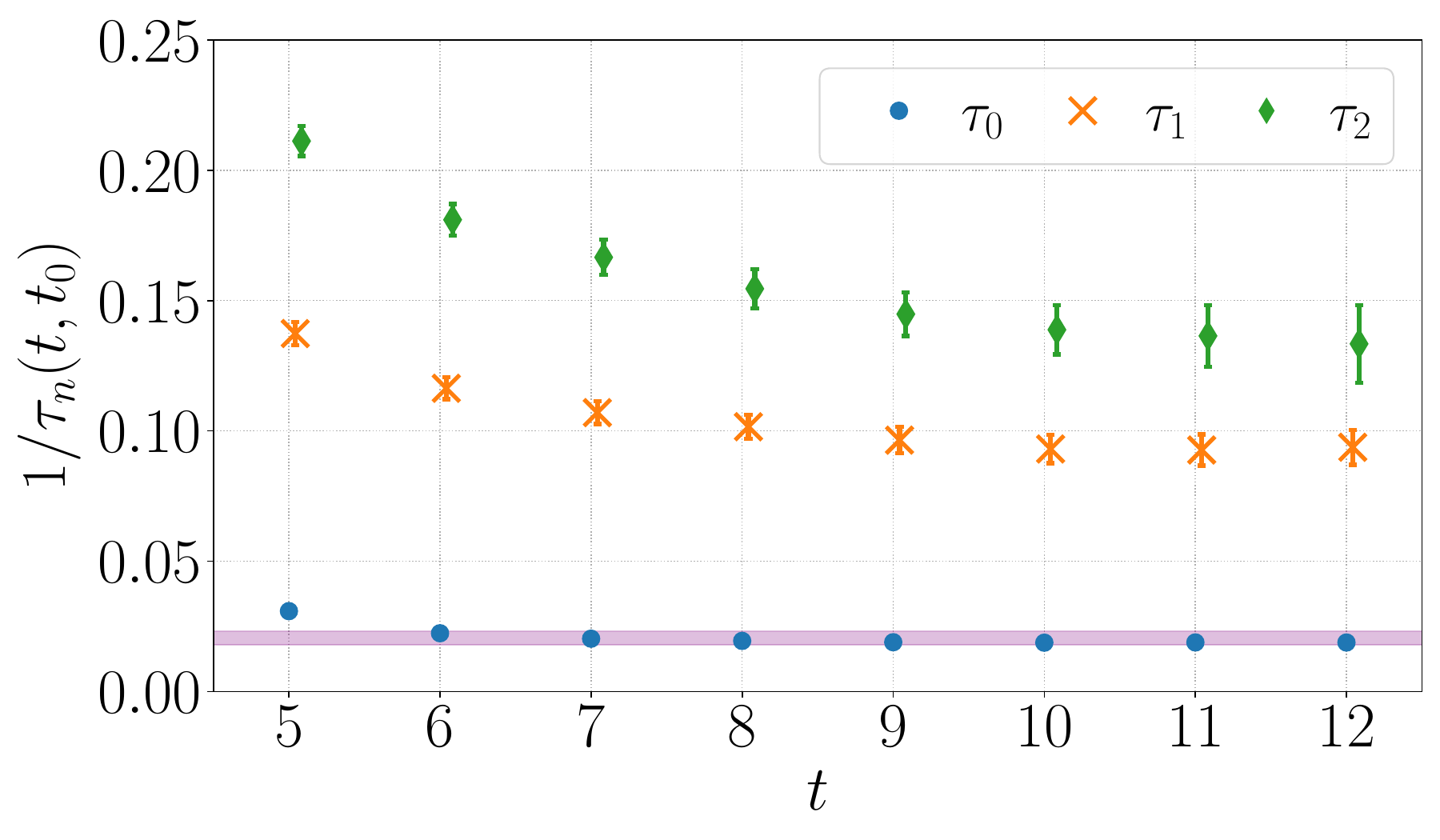}
    \includegraphics[width=.45\textwidth,keepaspectratio]{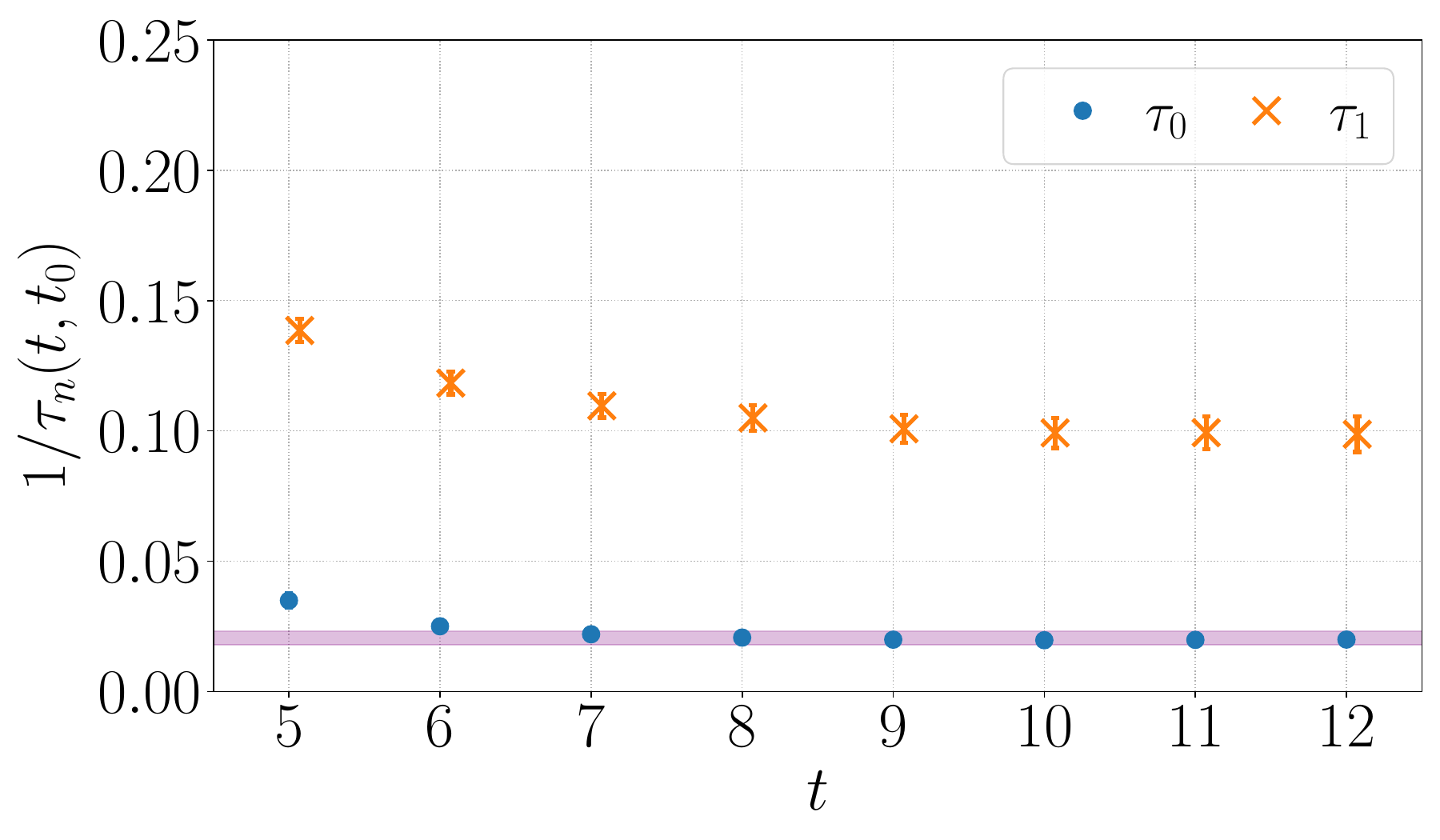}
    \caption{Autocorrelation modes obtained from the GEVP and the pGEVP applied to the MC autocorrelation function of the observables $E$, $\chi_\mathrm{M}$, $\chi_\mathrm{T}$ and $\xi_\mathrm{G}$ in the $\CPN$ model, see subsection~\ref{subsec:cpn}. The purple horizontal band corresponds to $1/\tau_\mathrm{int}$ of $\chi_\mathrm{T}$, cf. Table~\ref{tab:CPN_obs_tau}. The modes $\tau_1$ and $\tau_2$ are slightly shifted along the $x$-axis for better readability. The extraction is performed by fixing $t-t_0=5$ in both cases. \textit{Left}: full GEVP. \textit{Right}: $2 \times 2$ pGEVP at $t_P = 4$.}
    \label{fig:GEVP_MC_CPN_comparison}
\end{figure}

The projection preserves the spectrum and the matrix elements of the original ``operators'' are obtained by replacing the numerator in Eq.~\eqref{eq:GEVP_coeffs} with $e^{E_n t} [\widetilde{v} \,\mathbb{P}^T \Gamma][\Gamma \mathbb{P} \widetilde{v}\,]$, where the $\widetilde{v}\,$'s are the eigenvectors of the pGEVP. A numerical comparison between the standard and projected GEVP is shown in Fig.~\ref{fig:GEVP_MC_CPN_comparison} using the same data discussed in the main text from the $\CPN$ model, namely the MC autocorrelation function involving four observables. The first three states extracted from the full $4\times 4$ GEVP are compared with the two states obtained from the projected $2 \times 2$ case. In Fig.~\ref{fig:GEVP_MC_CPN_comparison} we zoom in a region at short distances to highlight the correctness of the approach. The full $4\times 4$ matrix used for the diagonalization becomes unstable at $t\approx 13$, where the fourth (not shown) and third states become degenerate. On the contrary the pGEVP develops long plateaus to much longer $t$ coordinates.

\bibliography{biblio}
\bibliographystyle{apsrev4-1}

\end{document}